\documentclass[journal]{IEEEtran} 
\IEEEoverridecommandlockouts
\usepackage{graphicx}                                      
\usepackage{amssymb,amsthm,bm}  
\usepackage[cmex10]{amsmath} 
\usepackage{algorithm}
\usepackage{algorithmic}  

\usepackage{multirow}   
\usepackage{tikz}                   
\newtheorem{theorem}{Theorem}
\newtheorem{lemma}{Lemma}
\newtheorem{corollary}[lemma]{Corollary}
\newtheorem{remark}{Remark}

\newtheorem{proposition}{Proposition}

\newcommand{\utag}[2]{\mathop{#2}\limits^{\text{(#1)}}}
\newcommand{\uref}[1]{(#1)}

\long\def\symbolfootnote[#1]#2{\begingroup
\def\thefootnote{\fnsymbol{footnote}}\footnote[#1]{#2}\endgroup}

\usepackage{xcolor,colortbl}

\usepackage{mathrsfs}

\newcommand{\Ie}{i.e., }
\newcommand{\Eg}{e.g., }

\usepackage{diagbox}
\usepackage{color}

\usepackage{tikz}
\usetikzlibrary{automata,positioning,chains,fit,shapes,calc}
\usetikzlibrary{arrows}
\usepackage{caption}
\usepackage{booktabs}
\usepackage{cite}
\usepackage{url}

\usepackage{subcaption}
\usepackage{cases}

\makeatletter
\newcommand{\mathleft}{\@fleqntrue\@mathmargin0pt}
\newcommand{\mathcenter}{\@fleqnfalse}
\makeatother

\usepackage{pgfplots}
  \pgfplotsset{compat=newest}

  \usepgfplotslibrary{patchplots}
  \usepackage{grffile}
  \definecolor{darkgreen}{RGB}{34,139,34}
  \definecolor{lightblue}{rgb}{0.00000,0.44700,0.74100}
  \definecolor{mycolor3}{rgb}{0.92900,0.69400,0.12500}%
  
  \usepackage{mathtools}

\usepackage{bbm}
\usepackage{tabu}
\usepackage{soul}

\newcommand{\pb}[1]{p\left(#1\right)}
\newcommand{\pbb}[1]{\mathbb{P}\left\{#1\right\}}

\usepackage{tikz}
\usetikzlibrary{er,automata,positioning,chains,fit,shapes,calc,arrows,plotmarks,arrows.meta}
\usepackage{caption}
\usepackage{varwidth}
\usetikzlibrary{calc}

\usepackage{transparent}

\begin{document}

\title{Intermittent Private Information Retrieval with Application to Location Privacy}

\author{
    Fangwei~Ye,~\IEEEmembership{Member,~IEEE},
     and Salim~El~Rouayheb,~\IEEEmembership{Senior~Member,~IEEE}

\thanks{
F.~Ye was with the Department of Electrical and Computer Engineering, Rutgers, The State University of New Jersey, Piscataway, NJ 08854, USA. He is with the Broad Institute of MIT and Harvard, Cambridge, MA 02142, USA (email: fye@broadinstitute.org).

S.~El~Rouayheb is with the Department of Electrical and Computer Engineering, Rutgers, The State University of New Jersey, Piscataway, NJ 08854, USA (email: salim.elrouayheb@rutgers.edu).
}

\thanks{
    This work of S.~El~Rouayheb was supported in part by NSF Grant CCF 1817635.
}

}

\maketitle

\begin{abstract}
We study the problem of intermittent private information retrieval with multiple servers, in which a user consecutively requests one of $K$ messages from $N$ replicated databases such that part of requests need to be protected while others do not need privacy. Motivated by the location privacy application, the correlation between requests is modeled by a Markov chain. We propose an intermittent private information retrieval scheme that concatenates an obfuscation scheme and a private information retrieval scheme for the time period when privacy is not needed, to prevent leakage incurred by the correlation over time. In the end, we illustrate how the proposed scheme for the problem of intermittent private information retrieval with Markov structure correlation can be applied to design a location privacy protection mechanism in the location privacy problem. 
\end{abstract}


\section{Introduction}
\label{introduction}
Privacy-preserving mechanism \cite{sweeney2002k,dwork2008differential,bezzi2010information} has been intensively studied because of the upsurge in privacy concerns. An emerging application is  location privacy, since location-based service becomes an integral part of daily life. Location privacy has attracted significant attention recently \cite{shokri2010unraveling,gedik2007protecting,hua2017geo,chatzikokolakis2014predictive,xiao2015protecting,shokri2011quantifying,shokri2016privacy,Tandon2019location,Gunduz2021location,oya2017back,bindschaedler2016synthesizing,shokri2012protecting}, and particularly from an information-theoretic perspective \cite{Tandon2019location,Gunduz2021location}. However, it has not been fully addressed that how the correlation between locations corrodes privacy, especially when a user may not need privacy all the time due to the overheads incurred by the privacy-preserving mechanism. To capture the impact of correlation in location privacy systematically, we formulate a theoretical problem in the framework of information-theoretic private information retrieval (PIR) \cite{chor1995private,Sun2017capacity}. In particular, we are interested in the Markov structure correlation, as a commonly adopted mobility model that models the correlation between locations is the Markov model \cite{shokri2011quantifying,Tandon2019location,Gunduz2021location,shokri2016privacy,chatzikokolakis2014predictive,xiao2015protecting}.

Private information retrieval (PIR) \cite{chor1995private,Sun2017capacity} has attracted significant attention recently due to its key role in understanding privacy in downloading scenarios. The PIR capacity, that is, the utility metric to measure download cost from databases, was characterized by Sun and Jafar \cite{Sun2017capacity}, in which the canonical setting is that a user is interested in retrieving one of the $K$ messages from $N$ replicated database while hiding the identity of the desired message. Many variants of the ordinary PIR problem have been studied in \cite{Sun2018symmetric,Sun2017colluding,Salim2018MDS,Salim2019sideinfo,Ulukus2018coded,Ulukus2018Byzantine,Ulukus2019privatesideinfo,Tandon2020PIR,Tian2019messagesize,Tian2020storage,li2018single,li2020single,li2020converse,Zhou2020MDScost,Wang2018eavesdrop}.

The new variation to be studied in this paper, namely intermittent private information retrieval, is motivated by the fact that privacy usually comes at a cost so a user may not need privacy all the time. Privacy-preserving mechanisms typically incur higher overheads in terms of computation, memory, and delay, etc. These incurred burdens may motivate the user to choose whether he/she needs privacy or not at certain times. For example, people may 
switch between normal and incognito modes in browsers depending on network connection and sensitivity of contents, etc. 

Under the intermittent PIR setting, when a user needs privacy, he/she has to use a PIR scheme. The question is what should be done when the user does not need privacy. One natural answer is a straightforward scheme, \Ie a scheme without any concern of privacy, which suffers from the fact that the user's behavior is usually correlated over time and hence a careless downloading at the current time will leak information about the request at the time instance that needs privacy. Another natural answer is a PIR scheme, which surely preserves privacy due to the one-shot nature of the PIR scheme~\cite{Sun2017capacity}. However, this conservative strategy generally sacrifices the efficiency, \Ie increasing the download cost, since it over-protects a request that does not need privacy. 

In this paper, we study the problem of intermittent private information retrieval with Markov structure correlation, in which a user consecutively requests one of $K$ time-varying messages from $N$ replicated databases at each time, such that part of requests (at some time periods) need privacy while others do not need privacy. The requests over time are correlated and we model the correlation by a first-order Markov chain as said.

We propose a solution that can be considered as a concatenation of an obfuscation scheme and a PIR scheme. In particular, the scheme can be viewed as a PIR scheme over a randomized subset of messages, where the subset is optimized according to the given Markov structure correlation between requests. Also, we bound the download cost of the concatenation scheme. 
The obfuscation scheme that optimizes the randomly chosen subsets first appeared as a primitive component in the ON-OFF privacy problem \cite{ONOFF-arxiv,ONOFF-TIFS} proposed by the authors, where the ON-OFF privacy problem can be regarded as an intermittent PIR problem with a \emph{single server} in the language of this paper. Therefore, the proposed scheme in this paper can be considered as extending the obfuscation scheme for the intermittent PIR with a single server therein to the setting of the intermittent PIR with \emph{multiple servers}.

To echo the location privacy motivation at the beginning, we will illustrate how the proposed scheme for the problem of intermittent PIR with Markov structure correlation can be applied to design a location privacy protection mechanism in the location privacy problem at the end. 



\emph{Organization:} 
The rest of the paper is organized as follows. In Section~\ref{sec:formulation}, we formulate the problem of intermittent private information retrieval with Markov structure correlation. The canonical case of two requests is discussed in Section~\ref{sec:two}, and the general case of a Markov chain is discussed in Section~\ref{sec:general}. In Section~\ref{sec:location}, we show how to apply the intermittent PIR scheme to the location privacy application. We conclude the paper in Section~\ref{sec:conclusion}. 


\emph{Notation:}
Throughout this paper, the probability distribution for a random variable $X$ that takes values in an alphabet $\mathcal{X}$ is denoted by $\{p_X(x): x \in \mathcal{X}\}$ with $p_X(x) = \pbb{X = x}$. When there is no ambiguity, $p_X(x)$ will be abbreviated as $p(x)$.

\section{Problem formulation
}
\label{sec:formulation}

We follow the terminology in \cite{Sun2017capacity} to introduce the setting of intermittent private information retrieval with multiple servers accompanying with the correlation between requests over time.

We assume that there are $N$ servers and $K$ \emph{time-varying} messages in the system. At each discrete time $t$, the messages $W_{1,t},\ldots,W_{K,t}$ are generated independently by $K$ information sources. At time $t$, each of the servers stores a replica of all $K$ updated messages $W_{1,t},\ldots,W_{K,t}$.
We slightly abuse the notations by dropping the time index $t$ from $W_{i,t}$ for notational simplicity, and the underlying time index $t$ will be clear in the context. Assume that $K$ messages (at each time) are mutually independent and each of the messages consists of $L$ independent bits that uniformly take values in the binary alphabet $\{0,1\}$. 

At each time $t$, the user is interested in retrieving a message from $\{W_{1},\ldots,W_K\}$. Let $\{X^{(t)}: t = 0, 1, 2, \ldots\}$ denote the requests, where each $X^{(t)}$ takes values in $[K]:=\{1,\ldots,K\}$. 
As mentioned in the introduction, the correlation model of requests is an essential attribute in the problem, and
 we are particularly interested in the case where the requests $X^{(t)}$ for $t = 0, 1, \ldots$ form a Markov chain.

The intermittence introduced in this paper is described as follows. The user may or may not wish to hide the identity of the message of interest at time $t$. 
Specifically, let $S^{(t)}$ denote the privacy status at time $t$, where $S^{(t)} = 1$ means that the user wishes to keep $X^{(t)}$ private and $S^{(t)} = 0$ means that the user is not concerned with privacy. 
We assume that the privacy status $S^{(t)}$ is completely chosen by the user, i.e., $S^{(t)}$ is viewed as a given parameter that is independent of the user's request $X^{(t)}$. 
We also assume that the privacy status $S^{(t)}$ is shared by both the servers and the user. In other words, we are not interested in hiding the privacy status in our formulation.  Without loss of generality, we assume that $S^{(0)} = 1$, i.e., $X^{(0)}$ needs privacy.

The same as the classical PIR setting, suppose that a user wants to retrieve a message $W_{X^{(t)}}$ at time $t$. 
To retrieve the message, the user generates $N$ queries $Q^{(t)}_1,\ldots,Q^{(t)}_N$ and the query $Q^{(t)}_i$ will be sent to the $i$-th server. To clarify, the user may generate the query for the request $X^{(t)}$ by utilizing all the \emph{causal} information, i.e., all the previous and the current requests $X^{(j)}$ for $j \leq t$, all the previous and the current privacy status $S^{(j)}$ for $j \leq t$, and all the previous queries $Q^{(j)}_i$ for $j < t$. More rigorously, the query at time $t$ is supposed to be generated by the query function that maps $\{X^{(j)}, S^{(j)}: j \leq t\}$, $Q^{(j)}_i$ for $j < t$ and some random key $\mathsf{F}^{(t)}$, to the query $Q^{(t)}_i$, i.e.,
\begin{equation}
\label{eq:query-encoding}
  \Phi_i: \{1,\ldots,K\}^{t+1}  \times \{0,1\}^{t+1}  \times \mathcal{Q}^t \times \mathcal{F} \rightarrow \mathcal{Q},
\end{equation}
where $\mathcal{Q}$ is supposed to be a common alphabet of queries for conciseness, and $\mathsf{F}^{(t)}$ denotes the random key\footnote{The key in this paper may be context-dependent, i.e., generated dependent of the input.} on the alphabet $\mathcal{F}$.

Upon receiving the a query $Q_i^{(t)}$, the $i$-th server  generates an answer $A_i^{(t)}$ to response to the query. We require that the answer $A_i^{(t)}$ is a deterministic function of the query $Q_i^{(t)}$ provided the stored messages. After receiving answers $A_1^{(t)},\ldots,A_N^{(t)}$, the user should be able to decode the desired message $W_{X^{(t)}}$ with zero error probability.

We would like to clarify two points about the setting. First, for any given privacy status, the query $Q_i^{(t)}$ may be viewed as a stochastic function of all the causal requests. Second, the messages are assumed to be time-varying, more precisely independent over time, so the answer $A_{i}^{(t)}$ only depends on the current messages and the query $Q_i^{(t)}$.

As said, the user should be able to decode the message of interest, which is referred to as \emph{correctness} requirement \cite{Sun2017capacity}. The correctness requirement is defined in the same way in this paper, \Ie
\begin{equation}
\label{eq:correctness}
  \begin{aligned}
       H(W_{X^{(t)}}|X^{(t)}, \mathsf{F}^{(t)}, Q_{1:N}^{(t)}, A_{1:N}^{(t)})  = 0, 
  \end{aligned}
\end{equation} 
where $Q_{1:N}^{(t)}:=\{Q_{i}^{(t)}:i=1,\ldots,N\}$ and $A_{1:N}^{(t)}:=\{A_{i}^{(t)}:i=1,\ldots,N\}$.

The other requirement of the system is the \emph{privacy} requirement. For our intermittent PIR setting, we require that for any time $t$, given all previous queries received by the $i$-th server, 
the query $Q_i^{(t)}$ should not reveal any information about the causal requests that need privacy, i.e.,
\begin{equation}
\label{eq:privacy}
  \begin{aligned}
     \text{[Privacy]} & & I(X^{(\mathcal{P}_{t})}; Q_{i}^{(t)}| Q_{i}^{(1)}, \ldots, Q_{i}^{(t-1)})  = 0, 
  \end{aligned}
\end{equation} 
for $i \in [N]$, where $X^{(\mathcal{P}_{t})}: = \{X^{(j)}: j \in \mathcal{P}_{t}\}$ and 
\begin{equation}
\label{eq:define-Pt}
  \mathcal{P}_{t} := \{j: S^{(j)} =1, j \leq t\}.
\end{equation} 
Note that $0 \in \mathcal{P}_{t}$ for any $t$ from the assumption $S^{(0)} = 1$.

The conditioning in \eqref{eq:privacy} serves to ensure causality by design. Barring this conditioning, privacy could be alternatively defined by
\begin{equation}
\label{eq:privacy-alter}
  \begin{aligned}
      I(X^{(\mathcal{P}_{t})}; Q_{i}^{(1)}, \ldots, Q_{i}^{(t)})  = 0. 
  \end{aligned}
\end{equation} 
However, this alternative definition implies that at some point $j < t$, the query $Q_{i}^{(j)}$ may be required to protect some \emph{future} request $X^{(j')}$ such that $j'>j$ and $j' \in \mathcal{P}_{t}$, i.e.,
\[I(X^{(j')}; Q_{i}^{(j)}) = 0, \]
induced by \eqref{eq:privacy-alter}. This generally means that the adversary may attempt to infer future requests. Given the correlation between the requests and the assumption that the user does not know (or infer) the future requests in advance (adhere to causality in our formulation), the alternative privacy definition enforces a less interesting solution that the query $Q_{i}^{(j)}$ must be independent of $X^{(j)}$, i.e., a standard PIR scheme all the time. For this reason, we adopt \eqref{eq:privacy} as the privacy requirement, which leads to a theoretically interesting and meaningful problem.

Conventionally, the utility metric is defined by the normalized download cost. Let $\ell_i^{(t)}$ denote the length of the answer $A_{i}^{(t)}$, and the normalized download cost of the $i$-th server is given by
\begin{equation*}
  \alpha^{(t)}_i := \frac{\mathbb{E}[\ell_i^{(t)}]}{L},
\end{equation*}
that is the expected amount of downloaded data per bit desired message from the $i$-th server. Correspondingly, the total download cost is 
\begin{equation*}
  \alpha^{(t)} = \sum_{i=1}^{N} \alpha^{(t)}_i.  
\end{equation*} 
Clearly, we are aimed to minimize the total download cost.

\section{Canonical Case: two requests}
\label{sec:two}
In this section, we start from the canonical case of two requests, which is the first step to understand the impact of the correlation in the intermittent private information retrieval problem. Also, we will see later it indeed serves as the key component to solve the general problem where the requests are modeled by a Markov chain.

Let $X^{(0)}$ and $X^{(1)}$ be two random variable taking values in $[K]$, representing two requests at time $t = 0$ and $t = 1$ respectively. 
Suppose that $S^{(0)} = 1$ and $S^{(1)} = 0$, i.e.,  $X^{(0)}$ at time $t=0$ is a request that needs privacy while $X^{(1)}$ at time $t=1$ is a request that does not need privacy. 
The initial probability distribution of $X^{(0)}$ is denoted by $\pi_0$, and the transition probabilities $p(x^{(1)}|x^{(0)})$ for $x^{(0)}, x^{(1)} \in [K]$ are known.

By invoking the privacy requirement in \eqref{eq:privacy}, the designed queries $Q_i^{(0)}$ and $Q_i^{(1)}$ for $i \in [N]$ at two time periods $t = 0$ and $t = 1$ should satisfy
\begin{equation}
\label{eq:privacy-0}
      I(X^{(0)}; Q_{i}^{(0)}) = 0, 
\end{equation}
and
\begin{equation}
\label{eq:privacy-1}
     I(X^{(0)}; Q_{i}^{(1)}| Q_{i}^{(0)})  = 0. 
\end{equation}
Note that $\mathcal{P}_0 = \mathcal{P}_1 = \{0\}$ provided the privacy status $S^{(0)} = 1$ and $S^{(1)} = 0$.

First, we notice that the sub-problem of minimizing the download cost $\alpha^{(0)}$ at $t = 0$, satisfying \eqref{eq:privacy-0}, is exactly a original PIR problem, and the minimum download cost is known in \cite{Sun2017capacity}, i.e., $\min \alpha^{(0)} = C(N,K)$, where
\begin{equation}
\label{eq:PIR-capacity}
   C(N,K): = 1+N^{-1}+N^{-2}+ \cdots + N^{-K+1},   
\end{equation} 
which can be achieved by the PIR-capacity achieving scheme therein. 

Therefore, the interesting part is to ask if there exists a better retrieval mechanism at $t = 1$ when the privacy is not needed, while preserving the privacy of $X^{(0)}$.

Provided that $Q^{(0)}_i$ is the query of the PIR capacity-achieving scheme, i.e., 
\begin{equation*}
    I(X^{(0)}; Q_{i}^{(0)})   = 0 ~~ \text{and} ~~ 
    I(Q_{i}^{(1)}; Q_{i}^{(0)}|X^{(0)}) = 0,
\end{equation*}
where the latter one follows because the query $Q^{(0)}_i$ of a PIR scheme only depends on $X^{(0)}$ and some random key, the privacy requirement \eqref{eq:privacy-1} can then be written by 
\begin{equation}
\label{eq:privacy-two-1}
  \begin{aligned}
     I(X^{(0)}; Q_{i}^{(1)})  = 0, ~~\forall i \in [N],
  \end{aligned}
\end{equation} 
which is formally stated in the following proposition and the justification is deferred to the appendix.
\begin{proposition}
\label{proposition:intro}
  For any $i \in [N]$, given $I(X^{(0)}; Q_{i}^{(0)}) = 0$ and $I(Q_{i}^{(1)}; Q_{i}^{(0)}|X^{(0)}) = 0$, we know that $$I(X^{(0)};Q_{i}^{(1)}|Q_{i}^{(0)}) = 0$$ if and only if $$I(X^{(0)}; Q_{i}^{(1)}) = 0.$$
\end{proposition}

Therefore, we will focus on designing queries $Q_i^{(1)}$, $i \in [N]$ satisfying the requirement \eqref{eq:privacy-two-1} in the sequel. We start by introducing some necessary notations and stating the result.

Suppose the transition probabilities
$p(x^{(1)}|x^{(0)})$ for $x^{(0)},x^{(1)} \in [K]$ are given. For any $i \in [K]$, suppose that 
\begin{equation}
\label{eq:ordering}
\begin{aligned}
  & \pbb{X^{(1)} = i|X^{(0)}= v_{i,1}} \leq \cdots  \leq \pbb{X^{(1)} = i|X^{(0)}=v_{i,K}},
\end{aligned}
\end{equation}
\Ie ordering the likelihood probabilities of $\pb{x^{(1)}|x^{(0)}}$, where $v_{i,1},\ldots,v_{i,K}$ are $K$ distinct elements in $[K]$. Let $\lambda_j$ be the summation of the $j$-th minimal likelihood probabilities for each possible value of $X^{(1)}$, \Ie 
\begin{equation}
\label{eq:lambda}
  \lambda_j : = \sum_{i \in [K]} \pbb{X^{(1)}=i|X^{(0)}=v_{i,j}},\ j=1,\ldots,K.
\end{equation}
Also, let 
\begin{equation}
\label{eq:sigma}
  \sigma : = \max\{j:\lambda_j \leq 1\}, 
\end{equation}
and 
\begin{equation}
\label{eq:differ}
  \theta_j : = \min\{1,\lambda_j\} - \min\{1,\lambda_{j-1}\},
\end{equation}
i.e., $\lambda_{j}- \lambda_{j-1}$ for $j \leq \sigma$, $1- \lambda_{\sigma}$ for $j = \sigma+1$, and $0$ for $j > \sigma+1$. All these parameters can be obtained from the given transition probabilities
$p(x^{(1)}|x^{(0)})$ for $x^{(0)},x^{(1)} \in [K]$.


\begin{theorem}
\label{thm:theorem}
    For any given transition probabilities $p(x^{(1)}|x^{(0)})$ for  $x^{(0)},x^{(1)} \in [K]$, there exists an intermittent private information retrieval scheme with download cost 
    \begin{equation}
    \label{eq:theorem}
      \alpha^{(1)} = \mathbb{E}\left[\left(1- \frac{1}{N} \right)^{-1}\left(1- \frac{1}{N^{|U|}} \right)\right],
    \end{equation} 
    for some random variable $U$ that takes value in the power set of $[K]$ such that 
      \begin{equation}
      \label{eq:theorem-U}
        \pbb{|U| \leq i} \geq \sum_{j=1}^{i} \theta_j,~ \forall\, i=1,\ldots,K.
      \end{equation}
\end{theorem}

To clarify, Theorem~\ref{thm:theorem} states that there exists some $U$ satisfying~\eqref{eq:theorem-U}, such that the download cost of the scheme at time $t = 1$ is $\mathbb{E}[C(N,|U|)]$, as shown in \eqref{eq:theorem}. In fact, the auxiliary random variable $U$ represents an obfuscation scheme in our design. Therefore, \eqref{eq:theorem} implies that the download cost of our intermittent private information retrieval scheme depends on the design of the obfuscation scheme, and \eqref{eq:theorem-U} guarantees that there exists an obfuscation scheme satisfying \eqref{eq:theorem-U}.

As $C(N,|U|)$, i.e., the expression with the expectation in \eqref{eq:theorem}, is increasing with $|U|$ for a given $N$, it suggests that if the probability of $U$ of small size is larger, then the download cost is generally smaller. 
However, due to the privacy requirement, it may not be possible to make $|U|$ too small, \Eg the extreme case is that $|U|=1$ with probability $1$. 
Nevertheless, \eqref{eq:theorem-U} guarantees the existence of a random variable $U$ such that the distribution of $U$ satisfies \eqref{eq:theorem-U}, where the worst case is that
  \begin{equation}
  \label{eq:theorem-worst-case-U}
    \pbb{|U| = i} =  \theta_i, \ i = 1,\ldots, K.
  \end{equation}
As such, we have the following corollary immediately from the theorem.
\begin{corollary}
\label{thm:corollary}
    For any given transition probabilities $p(x^{(1)}|x^{(0)})$ for  $x^{(0)},x^{(1)} \in [K]$, there exists an intermittent private information retrieval scheme with download cost 
       \begin{equation}
    \label{eq:theorem-worst-case}
      \alpha^{(1)} \leq \sum_{i=1}^{K} \theta_i \left(1- \frac{1}{N} \right)^{-1}\left(1- \frac{1}{N^{i}} \right).
    \end{equation} 
\end{corollary}
It is clear that the corollary can be established by showing that \eqref{eq:theorem-worst-case-U} is indeed the worst case of \eqref{eq:theorem-U}, in terms of the corresponding download cost, which is justified  as follows:
\begin{align*}
    \mathbb{E}\left[C(N,|U|)\right] 
    & = \sum_{i=1}^K   C(N,i) \, \pbb{|U| = i} \\
    & = \sum_{i=1}^K   \left(C(N,i) - C(N,i-1)\right) \sum_{j = i}^{K} \pbb{|U| = j} \\
    & = \sum_{i=1}^K   \left(C(N,i) - C(N,i-1)\right)  \pbb{|U| \geq i}.
\end{align*}
Since \eqref{eq:theorem-U} implies that 
\[\pbb{|U| \geq i} \leq \sum_{j = i}^{K} \theta_j \]
by the fact that $\sum_{j=1}^{K} \theta_j = 1$ from the definition of $\theta_j$\,(c.f.\eqref{eq:differ}), and $C(N,i)$ is increasing with $i$, we immediately obtain that
\begin{align*}
    \mathbb{E}\left[C(N,|U|)\right] 
    & = \sum_{i=1}^K   \left(C(N,i) - C(N,i-1)\right)  \pbb{|U| \geq i} \\
    & \leq \sum_{i=1}^K   \left(C(N,i) - C(N,i-1)\right)  \sum_{j = i}^{K} \theta_j \\
    & = \sum_{i=1}^K   C(N,i) \, \theta_i,
\end{align*} 
which completes the justification.

Another immediate observation of the theorem is that the right-hand side of~\eqref{eq:theorem} is exactly the same as the inverse of the PIR capacity in \cite{Sun2017capacity}, when $|U| = K$ certainly, i.e., $U = [K]$ with probability $1$. As said, the expression within the expectation is increasing with $|U|$ and a trivial upper bound on $|U|$ is $K$, which implies that the download cost specified by \eqref{eq:theorem} and \eqref{eq:theorem-U} is always better than the download cost of a standard PIR scheme in general.


\subsubsection*{Example}
To better illustrate the impact of correlation, we present an example here to show the relation between the download cost $\alpha^{(1)}$ and the given transition probabilities $\pb{x^{(1)}|x^{(0)}}$. We study the simplest case $N = K = 2$, and we write $\pb{x^{(1)}|x^{(0)}}$ explicitly by 
the probability transition matrix
\begin{equation*}
 P =
\begin{bmatrix}
       1- \alpha &  \alpha         \\
       \beta  & 1- \beta 
\end{bmatrix},
\end{equation*}   
such that $0 \leq \alpha, \beta \leq 1$. $P_{i,j}$ denotes $\pbb{X^{(1)} = j |X^{(0)} = i}$.
By inspecting the definition \eqref{eq:differ}, we know that 
\[\theta_1 = \min \{\alpha + \beta, 2 - \alpha - \beta\}, \ \theta_2 = 1 - \theta_1.\]

Without loss of generality, we assume that $\alpha + \beta \leq 1$. 
By \eqref{eq:theorem-worst-case}, we have 
\begin{equation*}
  \alpha^{(1)} \leq \frac{3}{2} - \frac{1}{2}(\alpha + \beta),
\end{equation*} 
where $3/2$ is the download cost of a PIR scheme over $K = 2$ messages for $N = 2$ servers and $\alpha + \beta$ somehow represents the correlation. We can clearly see two extreme cases. If $\alpha + \beta = 1$, i.e., requests are independent, then $\alpha^{(1)} = 1$, which implies that we can retrieve the desired message directly (or view it as a PIR scheme for $1$ message and $2$ servers). If $\alpha + \beta = 0$, i.e., requests are deterministic by each other, then $\alpha^{(1)} = 3/2$, which corresponds to the download cost of a PIR scheme for $K = N =2$.

In the next section, we will describe a scheme achieving the download cost shown in Theorem~\ref{thm:theorem}. In particular, we will show there exists some $U$ (obfuscation scheme) satisfying \eqref{eq:theorem-U}, such that the average download cost of the resulting intermittent private information retrieval scheme is $\mathbb{E}\left[C(N,|U|)\right]$.

\subsection{Concatenation Scheme}
\label{sec:formulation-scheme}
We consider a concatenation of an obfuscation scheme and a standard PIR capacity-achieving\footnote{Any capacity-achieving PIR scheme works, and we choose the pioneering one \cite{Sun2017capacity} for concreteness.} scheme \cite{Sun2017capacity}, to achieve the download cost as shown in Theorem~\ref{thm:theorem}.

A helpful observation on PIR capacity \cite{Sun2017capacity} is that the capacity is decreasing with the number of messages, so the general idea here is that we randomly choose a subset $U \subset [K]$ of messages, and implement the PIR scheme over the selected subset of messages. 
Generally speaking, the download cost is smaller when the size of the subset is smaller. However, privacy may not hold when the size of the subset is too small. For example, if $|U| = 1$ certainly, \Ie only downloading the desired message, the privacy may be broken since the server immediately knows which message is being retrieved.
 On the other hand, if $|U| = K$ certainly, \Ie always using a standard PIR scheme over $K$ messages, the privacy holds but with a high download cost. Therefore, we have to optimize the randomized way of choosing such a subset $U$ to reduce its size while preserving privacy.

More precisely, we first obfuscate the request $X^{(1)}$ to a set $U \subseteq [K]$ that includes $X^{(1)}$, and then retrieve the message $W_{X^{(1)}}$ privately by taking the PIR capacity-achieving scheme over a subset of messages $\{W_i: i \in U\}$. In this way, the PIR scheme preserves the identity of $X^{(1)}$ provided $U$, \Ie only information about $U$ is leaked, and the obfuscation is designed to guarantee that no information about $X^{(0)}$ can be obtained from $U$.

\subsubsection*{Example}
Before describing the scheme in details, we study the simplest example $N=K=2$ to illustrate the idea. The setting of servers is the same as the example in \cite{Sun2017capacity}, \Ie each server stores a full copy two messages (at time $t=1$) $(a_1,a_2,a_3,a_4)$ and $(b_1,b_2,b_3,b_4)$.

Suppose that the joint distribution of $X^{(0)}$ and $X^{(1)}$ is given in Table~\ref{table:example-1}. The obfuscation set $U$ can be designed according to the conditional probabilities in Table~\ref{table:example-2}.

\begin{table}[htbp]
\centering
\begin{tabular}{|c|c|c|}
\hline 
\diagbox{$X^{(0)}$}{$X^{(1)}$} & $1$ & $2$  \\
\hline 
$1$ & $3/8$ & $1/8$  \\
\hline
$2$ & $1/8$ & $3/8$ \\
\hline                                          
\end{tabular}
\caption{Joint probability distribution $\pb{x^{(0)},x^{(1)}}$.}
\label{table:example-1}
\end{table}

\begin{table}[htbp]
\centering
\begin{tabular}{|c|c|c|c|}
\hline 
\diagbox{$(X^{(0)},X^{(1)})$}{$U$} & $\{1\}$ & $\{2\}$ & $\{1,2\}$  \\
\hline 
$(1,1)$ & $1/3$ & $0$ & $2/3$\\
\hline
$(1,2)$ & $0$ & $1$ & $0$\\
\hline                    
$(2,1)$ & $1$ & $0$ & $0$\\
\hline
$(2,2)$ & $0$ & $1/3$ & $2/3$\\
\hline                      
\end{tabular}
\caption{Conditional probabilities $\pb{u|x^{(0)},x^{(1)}}$.}
\label{table:example-2}
\end{table}

\begin{table}[t]
\centering
\begin{tabular}{|c|c|c|}
\hline 
 Probability & \text{DB1} & \text{DB2}   \\
\hline 
\multirow{2}{*}{$\frac{2}{3}$} & $a_1,b_1$ & $a_2,b_2$ \\
    & $a_3+b_2$ & $a_4+b_1$\\
\hline 
\multirow{2}{*}{$\frac{1}{3}$} & $a_1$ & $a_3$ \\
    & $a_2$ & $a_4$\\
\hline                      
\end{tabular}
\caption{Time-sharing of two schemes for $X^{(0)} = X^{(1)}=1$ based on $U$.}
\end{table}

Assume that $X^{(0)} = X^{(1)}=1$. With probability $\frac{2}{3}$, the user will request the first message via a standard $N=K=2$ PIR scheme, \Eg querying for $(a_1,b_1,a_3+b_2)$ from the first server and $(a_2,b_2,a_4+b_1)$ from the second server, \Ie totally $6$ bits downloaded for a message of $4$ bits. With probability $\frac{1}{3}$, the user will directly request the message $1$ as desired, \Eg directly querying for $(a_1,a_2)$ from the first server and $(a_3,a_4)$ from the second server. We can check that $\frac{16}{3}$ bits are downloaded on average to retrieve the first message when $X^{(0)} = X^{(1)} = 1$.

Similarly, if $X^{(0)} = 2$ and $X^{(1)}=1$, the user will directly request the message $1$ as desired certainly from Table~\ref{table:example-2}, \Eg querying for $(a_1,a_2)$ from the first server and $(a_3,a_4)$ from the second server.  

When one of the servers, e.g., the first server,  receives the queries for $(a_1,a_2)$, although it immediately knows that the request at this time is $X^{(1)} = 1$, the privacy of $X^{(0)}$ is still preserved, since $X^{(0)} = 1$ and $X^{(0)} = 2$ are equally likely when $(a_1,a_2)$ is retrieved, i.e.,
\[
\begin{aligned}
  & \pbb{Q_1^{(1)} = (a_1,a_2) |X^{(0)} = 1} \\
  & ~~~~~~~~~~~~ = \pbb{Q_1^{(1)} = (a_1,a_2) |X^{(0)} = 2} = \frac{1}{4},
\end{aligned}\] 
due to the design of $U$ for the given correlation between $X^{(0)}$ and $X^{(1)}$.

Now, we describe the concatenation scheme in details as follows.

\emph{Obfuscation:} Suppose that $U$ is a subset of $[K]$, \Ie $U$ takes values in the power set of $[K]$, denoted by $\mathscr{P}_{K}$. Choose $U$ based on $X^{(1)}$ and $X^{(0)}$, more precisely the conditional probability $\pb{u|x^{(1)},x^{(0)}}$ for any given $p(x^{(1)},x^{(0)})$, to be a solution to the following ``optimization'' problem: 
\begin{equation}
\label{eq:achievable-LP-0}
    \begin{aligned}
    & \underset{U}{\text{minimize}}
    & & \mathbb{E}\left[C(N,|U|)\right]  & \\
    & \text{subject to}
    & & X^{(1)} \in U,   & \\
    & & & U~\text{is independent of}~X^{(0)}. & \\
    \end{aligned}
\end{equation}
Note that $U$ is a random variable and the expectation in the objective function is over $U$. A more standard formulation of this optimization problem, i.e., describing the decision variables explicitly, is deferred to the end of this section. Here, we keep this neat formulation to illustrate the basic idea of the obfuscation scheme. The constraint $X^{(1)} \in U$ represents that $X^{(1)} \in U$ certainly, or more precisely 
$p(u, x^{(1)}) = 0$ for $x^{(1)} \notin u$. 
The two constraints are indeed imposed to closely depict the intuitive idea of the scheme, i.e., obfuscating the request $X^{(1)}$ to a set $U$ that includes $X^{(1)}$ (necessary for the next PIR phase) and preserving the privacy of $X^{(0)}$. 
The discussion on solving this optimization problem is also deferred to the end of this section, and now let us just assume that the problem is solvable and the solution $p(u|x^{(1)},x^{(0)})$ can be obtained. After obtaining the solution $p(u|x^{(1)},x^{(0)})$, sample an obfuscation set $u$ according to $p(u|x^{(1)},x^{(0)})$ based on the observed requests $x^{(1)}$ and $x^{(0)}$.

\emph{Retrieval:} Given the request $X^{(1)}$ and the obfuscation set $U$, retrieve the message $W_{X^{(1)}}$ by using the standard PIR capacity-achieving scheme \cite{Sun2017capacity} for $|U|$ messages specified by $U$, \Ie constructing queries $Q_{i}^{(1)}$ for $i \in [N]$ from a PIR scheme with $N$ servers and $|U|$ messages.

Let us first examine the correctness and the privacy of this concatenated scheme. The correctness is an immediate consequence of the first constraint of \eqref{eq:achievable-LP-0}, since the retrieval scheme is just a private retrieval scheme to retrieve $W_{X^{(1)}}$ from $|U|$ messages including the desired message. 

For the privacy requirement, the obfuscation step constructs $U$ that is independent of $X^{(0)}$ as a constraint, so we have 
\begin{equation}
\label{eq:achievable-privacy-obfuscation}
     I(U;X^{(0)}) = 0.
\end{equation}   
Since the retrieval scheme is a standard PIR capacity-achieving scheme, we have
\begin{equation}
\label{eq:achievable-privacy-PIR}
   I(Q_{i}^{(1)};X^{(1)}|U) = 0,~\forall i \in [N],
\end{equation}
by examining the PIR scheme \cite{Sun2017capacity}, i.e., the query to an individual server $i$ does not leak any information about the request given the subset of messages that is of interest.

With \eqref{eq:achievable-privacy-obfuscation} and \eqref{eq:achievable-privacy-PIR}, we claim that
\[I(Q_{i}^{(1)};X^{(0)}) = 0,\]
which is the privacy requirement to be justified. 
Towards this end, consider
\begin{align*}
    I(Q_{i}^{(1)};X^{(0)}) & \leq I(Q_{i}^{(1)},U;X^{(0)}) \\
           & = I(U;X^{(0)}) + I(Q_{i}^{(1)};X^{(0)}|U) \\
           & \leq   I(U;X^{(0)}) + I(Q_{i}^{(1)}; X^{(0)}, X^{(1)} |U) \\
           & = I(Q_{i}^{(1)}; X^{(0)}, X^{(1)} |U), 
\end{align*} 
where $I(U;X^{(0)}) = 0$ follows from  \eqref{eq:achievable-privacy-obfuscation}. Since 
\begin{align*}
  & I(Q_{i}^{(1)}; X^{(0)}, X^{(1)} |U) \\
  & ~~~ =  I(Q_{i}^{(1)};X^{(1)}|U) + I(Q_{i}^{(1)};X^{(0)}|X^{(1)},U) \\
  & ~~~ =   0,
\end{align*}
where $I(Q_{i}^{(1)};X^{(1)}|U) = 0$ follows from \eqref{eq:achievable-privacy-PIR}, and $I(Q_{i}^{(X)};X^{(0)}|X^{(1)},U) = 0$ follows because $Q_{i}^{(1)}$ is only dependent of the random key given $X^{(1)}$ and $U$ for the private retrieval scheme, which implies 
\[X^{(0)} \rightarrow X^{(1)}, U \rightarrow Q_{i}^{(1)},\]
and thus $I(Q_{i}^{(1)};X^{(0)}|X^{(1)},U) = 0$. Hence, we can obtain that 
\[I(Q_{i}^{(1)};X^{(0)}) = 0,\]
that is to be proved.


Finally, let us evaluate the download cost of the scheme. Direct from the capacity result in \cite{Sun2017capacity}, the download cost for a given obfuscation set $U$ is 
\[C(N,|U|) = 1+N^{-1}+N^{-2}+ \cdots + N^{-|U|+1},\]
\Ie the (inverse) PIR capacity for $N$ servers and $|U|$ messages. Hence, the download cost of this concatenation scheme is given by 
\begin{equation*}
  \alpha^{(1)} = \mathbb{E}\left[C(N,|U|)\right] = \mathbb{E}\left[\left(1- \frac{1}{N} \right)^{-1}\left(1- \frac{1}{N^{|U|}} \right)\right],
\end{equation*} 
and the probability distribution of $U$ is specified by the solution to problem \eqref{eq:achievable-LP-0},  which indeed explains why we choose $\mathbb{E}\left[C(N,|U|)\right]$ as the objective function in \eqref{eq:achievable-LP-0}.

Therefore, we have justified that the concatenation scheme satisfies the correctness and the privacy requirements. The download cost is $\alpha^{(1)} = \mathbb{E}\left[C(N,|U|)\right]$, where $U$ represents an obfuscation scheme and can be \emph{any} feasible solution to the problem \eqref{eq:achievable-LP-0}. Referring to Theorem~\ref{thm:theorem}, the remaining part is to show the existence of some $U$ satisfying \eqref{eq:theorem-U}, i.e., the problem \eqref{eq:achievable-LP-0} has a feasible solution satisfying \eqref{eq:theorem-U}.

\subsection{
Existence of an admissible obfuscation
}

As said, the last step is show that there exists a solution to the problem \eqref{eq:achievable-LP-0} such that the resulting $U$ satisfies \eqref{eq:theorem-U}. Towards this end, we first interpret the problem \eqref{eq:achievable-LP-0} as a linear programming (LP), and write the decision variables and the objective function in a more explicit form.

The problem \eqref{eq:achievable-LP-0} can be viewed as a linear programming by treating each conditional probability $p(u|x^{(1)},x^{(0)})$ as a decision variable for any given $\pb{x^{(1)}, x^{(0)}}$, $x^{(0)}, x^{(1)} \in [K]$ and $u \in \mathscr{P}_{K}$. 
To see this, we first inspect the constraints. The first constraint $X^{(1)} \in U$ can be equivalently written by 
\begin{equation}
\label{eq:LP-decode}
  \pb{u|x^{(1)},x^{(0)}} = 0, \ \forall x^{(1)} \notin u.
\end{equation}

The second (independence) constraint can be written by
\begin{equation}
\label{eq:LP-privacy}
\begin{aligned}
  & \sum_{x^{(1)} \in [K]}  p(u|x^{(1)},x^{(0)})p(x^{(1)}|x^{(0)}) \\ 
  & ~~~~~ =  \sum_{x^{(1)} \in [K]} p(u|x^{(1)},\tilde{x}^{(0)})p(x^{(1)}|\tilde{x}^{(0)}),
\end{aligned}
\end{equation} 
for any $x^{(0)}, \tilde{x}^{(0)} \in [K]$ and $u \in \mathscr{P}_{K}$. For given $\pb{x^{(1)},x^{(0)}}$, both constraints are clearly linear with decision variables $p(u|x^{(1)},x^{(0)})$. 

Lastly, let us examine on the objective function. 
Although $C(N,|U|)$ seems a power function with $|U|$, $\mathbb{E}\left[C(N,|U|)\right]$ is indeed linear with decision variables $p(u|x^{(1)},x^{(0)})$, \Ie   
\begin{equation*} 
\begin{aligned}
  & \mathbb{E}\left[C(N,|U|)\right]  = \sum_{u \in \mathscr{P}_{K}} \pb{u} C(N,|u|) \\
  &  = \sum_{u \in \mathscr{P}_{K}} \sum_{x^{(0)},x^{(1)} \in [K]} p(x^{(0)},x^{(1)}) p(u|x^{(0)},x^{(1)}) C(N,|u|) \\
  &  =  \sum_{x^{(0)},x^{(1)}} p(x^{(0)},x^{(1)}) \sum_{c=1}^{K} C(N,c) \left( \sum_{u:|u|=c} p(u|x^{(0)},x^{(1)}) \right),  
\end{aligned}
\end{equation*} 
which is linear with $\pb{u|x^{(0)},x^{(1)}}$ for given $\pb{x^{(0)},x^{(1)}}$. 

By these interpretations, we write the optimization problem in a more explicit form:
\begin{equation}
\label{eq:achievable-LP}
    \begin{aligned}
    & \underset{p(u|x^{(1)},x^{(0)})}{\text{minimize}}
    & & \mathbb{E}\left[C(N,|U|)\right]   & \\
    & \text{subject to}
    & & \eqref{eq:LP-decode}, \eqref{eq:LP-privacy} & \\
    & & & \sum_{u \in \mathscr{P}_{K}} p(u|x^{(1)},x^{(0)}) = 1, \,\forall x^{(1)},x^{(0)},  & \\
        & & &  p(u|x^{(1)},x^{(0)}) \geq 0, \ \forall u, x^{(1)},x^{(0)}.  & \\
    \end{aligned}
\end{equation}
It is worth noting that the problem is always feasible since 
\[p(u|x^{(1)},x^{(0)}) = 
\begin{cases}
  1, & u=[K], \\
  0, & u \neq [K],
\end{cases}
\]
for $x^{(1)},x^{(0)} \in [K]$ is always a feasible solution for any given $p(x^{(1)},x^{(0)})$. In the context of PIR, it indicates that using the private retrieval scheme over $K$ messages is always feasible when querying for the request $X^{(1)}$.

It should be noted that a similar LP formulation was first discussed in \cite{ONOFF-arxiv} when authors studied a so-called ON-OFF privacy problem that can also be considered as the problem of intermittent PIR with a single server, although in a slightly different setting where two random variables $X^{(0)}$ and $X^{(1)}$ may not have the same alphabet therein.

Due to the exponential blowup in the number of decision variables and constraints, solving the LP instance numerically is generally intractable. Nevertheless, the following lemma, which is borrowed from \cite{ONOFF-arxiv} and interpreted with the notation in this paper, guarantees the existence of some solution that corresponds to \eqref{eq:theorem-U} as claimed in Theorem~\ref{thm:theorem}.

\begin{lemma}[{\cite[Lemma~3]{ONOFF-arxiv}}]
\label{lemma:achieve}
For any given random variables $X^{(0)}, X^{(1)} \in [K]$, there exists a random variable $U \in \mathscr{P}_K$ satisfying that $U$ is independent of $X^{(0)}$, $\pb{u|x^{(1)},x^{(0)}} = 0$ for $x^{(1)} \notin u$, and 
  \begin{equation}
  \label{eq:lemma-algorithm}
    \pbb{|U| \leq i} \geq \sum_{j=1}^{i} \theta_j,~ \forall\, i=1,\ldots,K.
  \end{equation}
\end{lemma}
In parlance of the optimization, Lemma~\ref{lemma:achieve} claims the existence of some feasible solution to the problem \eqref{eq:achievable-LP} (or equivalently \eqref{eq:achievable-LP-0}), for any given $X^{(1)}$ and $X^{(0)}$ (or $p(x^{(1)},x^{(0)})$), such that the resulting $U$ (or $\pb{u}$) satisfies the constraints in \eqref{eq:lemma-algorithm}. 
It is clear that \eqref{eq:lemma-algorithm} is exactly the same as \eqref{eq:theorem-U}, that is to be proved in Theorem~\ref{thm:theorem}.

The lemma is established by a constructive proof, i.e., constructing an admissible $\pb{u|x^{(1)},x^{(0)}}$ for $x^{(1)},x^{(0)} \in [K]$ and $u \in \mathscr{P}_K$, provided the given $\pb{x^{(1)}, x^{(0)}}$, or $\pb{x^{(1)}|x^{(0)}}$ (implying that the lemma holds for any initial probability distribution $\pi_0$). Instead of showing the detailed proof that can be found in \cite{ONOFF-arxiv}, we present an example to illustrate the basic idea of the construction, or roughly speaking the basic idea of finding some particular feasible solution to the optimization problem \eqref{eq:achievable-LP}.

\subsubsection*{Example}
Suppose that the transition probabilities $\pb{x^{(1)}|x^{(0)}}$ are given by
\begin{equation*}
  P =
  \begin{bmatrix}
    0.1 & 0.3 & 0.6 \\
    0.5 & 0.4 & 0.1 \\
    0.2 & 0.5 & 0.3
  \end{bmatrix},
\end{equation*}   
where $P_{i,j}=\pbb{X^{(1)}=j|X^{(0)}=i}$.

The designed probabilities $\pb{u,x^{(1)}|x^{(0)}}$ are represented in Table~\ref{table-example}, where the shaded cells of value $0$ come immediately from the condition $\pb{u,x^{(1)}|x^{(0)}} = 0$ for $x^{(1)} \notin u$. Throughout this example, we will show how to fill in the values of other cells. 

  \begin{table*}[t] 
        \centering
        \begin{tabular}{|c|c||c|c|c|c|c|c|c||c|}
            \hline
            \multicolumn{2}{|l||}{\diagbox[dir=SE,width=65pt,height=33pt]{\hspace{-1pt}$X^{(0)}$\hspace{5pt} $X^{(1)}$}{\vspace{-32pt}$U$}} 
            & $\{1\}$ & $\{2\}$ & $\{3\}$ & $\{1,2\}$ & $\{1,3\}$ & $\{2,3\}$ & $\{1,2,3\}$ & \cellcolor{yellow!20!white}\textcolor{black}{$P_{i,j}$}\\
            \hline 
            \multirow{3}{*}{$1$} & $1$ & $0.1$ & $0$ & $0$ & $0$ & $0$ & $0$ & $0$ &\cellcolor{yellow!20!white}$0.1$\\
            \cline{2-10} 
            & $2$ & \cellcolor{gray!30!white}$0$ & $0.3$ & \cellcolor{gray!30!white}$0$ & $0$ & \cellcolor{gray!30!white}$0$ & $0$ & $0$ &\cellcolor{yellow!20!white}$0.3$\\
            \cline{2-10} 
             & $3$ & \cellcolor{gray!30!white}$0$ & \cellcolor{gray!30!white}$0$ & $0.1$ & \cellcolor{gray!30!white}$0$ & $0.1+0.2$ & $0.1$ & $0.1$ &\cellcolor{yellow!20!white}$0.6$\\
            \hline 
            \multirow{3}{*}{$2$} & $1$ & $0.1$ & \cellcolor{gray!30!white}$0$ & $\cellcolor{gray!30!white}0$ & $0$ & $0.1+0.2$ & \cellcolor{gray!30!white}$0$ & $0.1$ & \cellcolor{yellow!20!white}$0.5$ \\
            \cline{2-10} 
             & $2$ & \cellcolor{gray!30!white}$0$ & $0.3$ & \cellcolor{gray!30!white}$0$ & $0$ & \cellcolor{gray!30!white}$0$ & $0.1$ & $0$ &\cellcolor{yellow!20!white}$0.4$\\
            \cline{2-10} 
             & $3$ & \cellcolor{gray!30!white}$0$ & \cellcolor{gray!30!white}$0$ & $0.1$ & \cellcolor{gray!30!white}$0$ & $0$ & $0$ & $0$ &\cellcolor{yellow!20!white}$0.1$\\
            \hline
            \multirow{3}{*}{$3$} & $1$ & $0.1$ & \cellcolor{gray!30!white}$0$ & \cellcolor{gray!30!white}$0$ & $0$ & $0.1$ & \cellcolor{gray!30!white}$0$ & $0$ &\cellcolor{yellow!20!white}$0.2$\\
            \cline{2-10} 
             & $2$ & \cellcolor{gray!30!white}$0$ & $0.3$ & \cellcolor{gray!30!white}$0$ & $0$ & \cellcolor{gray!30!white}$0$ & $0.1$ & $0.1$ &\cellcolor{yellow!20!white}$0.5$\\
            \cline{2-10} 
             & $3$ & \cellcolor{gray!30!white}$0$ & \cellcolor{gray!30!white}$0$ & $0.1$ & \cellcolor{gray!30!white}$0$ & $0.2$ & $0$ & $0$ &\cellcolor{yellow!20!white}$0.3$\\
             \toprule
      \end{tabular}
        \caption{The constructed $\pb{u,x^{(1)}|x^{(0)}}$ for the given $\pb{x^{(1)}|x^{(0)}}$.}
         \label{table-example}
    \end{table*}

\begin{itemize}
  \item $|U|=1$: For each $i \in [K]$, choose $U=\{i\}$, and let
  \begin{multline*}
    \pbb{U = \{i\},X^{(1)} = i|X^{(0)}= j} \\ = \pbb{X^{(1)}=i|X^{(0)} = v_{i,1}}
  \end{multline*}
  for all $j \in [K]$, i.e., $0.1$, $0.3$ and $0.1$ for $i = 1,2,3$, respectively, where $v_{i,1}$ is defined in \eqref{eq:ordering}.

  \item $|U|=2$: For each $i \in [K]$ and $v_{i,1}$, find a column index (of $P$) $c_{i}$ such that 
  \begin{multline*}
    \pbb{X^{(1)}=c_{i}|X^{(0)}=v_{i,1}} \\ \geq \pbb{X^{(1)}=c_{i}|X^{(0)}=v_{c_{i},2}} + \mu_{i},
  \end{multline*}
  where
   \begin{multline*}
   \mu_{i} = \pbb{X^{(1)}=i|X^{(0)}=v_{i,2}} \\ -  \pbb{X^{(1)}=i|X^{(0)}=v_{i,1}}.
  \end{multline*} 
  Choose $U=\{i,c_i\}$ and let
  \begin{equation}
  \label{eq:example-assign}
     \begin{aligned}
    & \pbb{U =\{i,c_i\},X^{(1)}=x|X^{(0)}=v_{i,j}} = \mu_i,
    \end{aligned}
  \end{equation}  
  for $j \geq 2$ and $x = i$ or $j < 2$ and $x = c_{i}$. As in this example, for $i=1$, we have $\mu_i = 0.1$, \Ie the second minimal value minus the minimum value in the first column of $P$, where $v_{i,1} = 1$ and $v_{i,2} = 3$. Let $c_i = 3$. Then we can check that 
  \[
  \begin{aligned}
     & 0.6 = \pbb{X^{(1)}=3|X^{(0)}=1} \\
     & ~~~~~~ \geq \pbb{X^{(1)}=3|X^{(0)}=v_{3,2}} + 0.1,
  \end{aligned}
 \]
  where $v_{3,2} = 2$ and hence $\pbb{X^{(1)}=3|X^{(0)}=v_{3,2}} = 0.1$. The process for $i=1$ finally configures the value $0.1$ for $U=\{1,3\}$ in the table. 

  This generally explains why we call it an obfuscation scheme. For each $i \in [K]$, we carefully find an index $c_i$ for $v_{i,1}$ and mix it with $i$ to form a set $U$ such that when observing $U$, there exists a pair $(x^{(1)},x^{(0)})$ generating $U$ for all $x^{(0)} \in [K]$. Note that since for different $i \in [K]$, the set $U$ may be the same, \Eg $U=\{1,3\}$ for both $i=1$ and $i=3$, so $\pb{u,x^{(1)}|x^{(0)}}$ is configured in an augmented way, \Ie the right-hand side of \eqref{eq:example-assign} is added to the left-hand side instead of being overwritten, such as $0.1+0.2$ in the cell.

  \item   $|U|=3$: Configure all remaining values constrained by $\pb{x^{(1)}|x^{(0)}}$, \Ie the summation of each row in the table. 
\end{itemize}

\begin{remark}
The general algorithm would basically extend the above process for $|U|=2$. Roughly speaking, for $|U| = c = 1,\ldots, \sigma$ and each $i \in [K]$, find an index $c_{i,j}$ for each $v_{i,j}$ such that $j \leq c-1$. Then choose $U=\{i, c_{i,j}: j \leq c -1 \}$ and configure 
  \begin{equation*}
     \begin{aligned}
    & \pbb{U,X^{(1)}=x|X^{(0)}=v_{i,j}} \\
    & = \pbb{X^{(1)}=i|X^{(0)}=v_{i,c}} -  \pbb{X^{(1)}=i|X^{(0)}=v_{i,c-1}},
    \end{aligned}
  \end{equation*}  
  for $j \geq c$ and $x = i$ or $j < c$ and $x = c_{i,j}$. It is worth noting that $c_{i,j}$ may be the same for different $j$, so the size of $U$ may be smaller than $c$. This observation indeed leverages the inequality \eqref{eq:lemma-algorithm} in the lemma, where the worst case is 
  \[\pbb{|U| = i} =  \theta_i,~ \forall\, i=1,\ldots,K,\]
  as mentioned. 
\end{remark}

\section{General Case: Markov chain}
\label{sec:general}
In this section, we will show how to use the two-requests scheme in Section~\ref{sec:formulation-scheme} as a building block to design an intermittent PIR scheme over time when the requests $X^{(t)}$, $t = 0,1, \ldots$ form a Markov chain. 

First, let 
\begin{equation}
\label{eq:define-tau}
  \tau(t) : = \max\{j: j \in \mathcal{P}_t\},
\end{equation}
where $\mathcal{P}_t$ is defined in \eqref{eq:define-Pt}. We may write $\tau(t)$ by $\tau$ for notational simplicity when the time index $t$ is clear in the context. Note that $\mathcal{P}_t$ is completely determined by the privacy status $S^{(t)}$, which is chosen by the user. Roughly speaking, $\tau(t)$ represents the latest time that the user needed privacy at time $t$.

Then the following proposition is a direct but useful consequence of the assumption of Markov structure correlation, and its proof is deferred to the appendix.
\begin{proposition}
\label{proposition:markov}
For any $i \in [N]$, if $Q_{i}^{(t)}$ is independent of $X^{(\tau(t))}$ conditioning on $ Q_{i}^{(0)}, \ldots, Q_{i}^{(t-1)}$, then $Q_{i}^{(t)}$ is independent of $X^{(\mathcal{P}_{t})}$ conditioning on $ Q_{i}^{(0)}, \ldots, Q_{i}^{(t-1)}$, i.e., 
\begin{equation}
\label{eq:privacy-latest}
  I(X^{(\tau(t))}; Q_{i}^{(t)}| Q_{i}^{(0)}, \ldots, Q_{i}^{(t-1)})  = 0, 
\end{equation}
implies that  
\begin{equation}
  I(X^{(\mathcal{P}_{t})}; Q_{i}^{(t)}| Q_{i}^{(0)}, \ldots, Q_{i}^{(t-1)})  = 0.
\end{equation} 
\end{proposition}

From Proposition~\ref{proposition:markov}, we know that it is sufficient to design queries $Q_{i}^{(t)}$ for $i \in [N]$ satisfying \eqref{eq:privacy-latest}, in order to guarantee the desired privacy \eqref{eq:privacy}. Roughly speaking, at time $t$, we need to design queries $Q_i^{(t)}, i \in [N]$ for the request $X^{(t)}$ while preserving the privacy of the request $X^{(\tau)}$.

Recall the scheme for the two-requests case in Section~\ref{sec:two}, where we design queries $Q_i^{(1)}, i \in [N]$ for the request $X^{(1)}$ while preserving the privacy of the request $X^{(0)}$. The roles of $X^{(0)}$ and $X^{(1)}$ are similar to $X^{(\tau)}$ and $X^{(t)}$ in this section, where the queries are designed for the retrieval purpose of the current request $X^{(t)}$ but they have to preserve privacy of some previous request $X^{(\tau)}$. 
Therefore, the scheme for the general Markov case is indeed similar to the canonical case of two requests that was discussed, where the main difference is that the prior distribution (e.g., $p(x^{(1)}|x^{(0)})$ in the previous section) as an input has to be updated at each time. In particular, let $U^{(t)}$ denote the obfuscation set at time $t$, and 
\begin{equation}
\label{eq:tracking-prob}
  p_{X^{(t)}|X^{(\tau)}}(x^{(t)}|x^{(\tau)},u^{(0)},\ldots,u^{(t-1)}),
\end{equation}
which serves the same role as $p(x^{(1)}|x^{(0)})$ in the two-requests case, has to be updated according to the generated $u^{(t)}$ at each time period.  For notational simplicity, let $U^{[0:t-1]} : = \{U^{(0)},\ldots,U^{(t-1)}\}$.

We summarize the above intuition by presenting a result that is similar to Theorem~\ref{thm:theorem}. Before that, we introduce a necessary notation that is similar to the one
defined in Section~\ref{sec:two}.
Let $\theta_j(u^{([0:t-1])})$ be defined the same as $\theta_j$ in \eqref{eq:differ}, but for the conditional probabilities $p_{X^{(t)}|X^{(\tau)}}(x^{(t)}|x^{(\tau)},u^{(0)},\ldots,u^{(t-1)})$ as shown in \eqref{eq:tracking-prob} for any given realizations $u^{([0:t-1])}$. Note that the random variables here are $X^{(t)}$ and $X^{(\tau)}$, serving the same roles as $X^{(1)}$ and $X^{(0)}$ in the definitions \eqref{eq:ordering}, \eqref{eq:lambda}, \eqref{eq:sigma} and \eqref{eq:differ}.

\begin{theorem}
Suppose that requests $X^{(t)}$ for $t = 0,1, \ldots$ form a Markov chain and the privacy status $S^{(t)}$ for $t = 0,1, \ldots$ are given. 
There exists an intermittent private information retrieval scheme with download cost 
    \begin{equation}
    \label{eq:theorem-t}
      \alpha^{(t)} = \mathbb{E}\left[\left(1- \frac{1}{N} \right)^{-1}\left(1- \frac{1}{N^{|U^{(t)}|}} \right)\right],
    \end{equation} 
    for some random variable $U^{(t)}$ that takes value in the power set of $[K]$ such that 
      \begin{equation}
      \label{eq:theorem-U-t}
        \pbb{|U^{(t)}| \leq i|U^{([0:t-1])} = u^{([0:t-1])}} \geq \sum_{j=1}^{i} \theta_j(u^{([0:t-1])}),
      \end{equation}
      for $i =1,\ldots,K$ and any $u^{([0:t-1])}$.
\end{theorem}
\begin{remark}
  Similar to Corollary~\ref{thm:corollary}, a slightly weaker but more explicit form of \eqref{eq:theorem-U-t} is that 
\begin{equation}
\label{eq:corollary-U-t}
        \pbb{|U^{(t)}| = j|U^{([0:t-1])} = u^{([0:t-1])}} = \theta_j(u^{([0:t-1])}),
\end{equation}
for $j =1,\ldots,K$ and any $u^{([0:t-1])}$.
\end{remark}

As said, the concatenation scheme that justifies the theorem can be described in the same manner as we did in the two-requests case, i.e.,
\begin{enumerate}
  \item Design an obfuscation set $U^{(t)}$ that includes $X^{(t)}$ and is independent of $X^{(\tau)}$ \emph{conditioning} on $U^{([0:t-1])}$, i.e.,  sampling an obfuscation set $u^{(t)}$ according to 
  \[p_{U^{(t)}|X^{(t)},X^{(\tau)}}(u^{(t)}|x^{(t)},x^{(\tau)}, u^{([0:t-1])}),\]
  that corresponds to an obfuscation scheme which is discussed right after the enumeration. 
  \item Query for the request $X^{(t)}$ by using a standard PIR scheme over messages specified by $U^{(t)}$.

\end{enumerate}

In particular, we modify the optimization problem \eqref{eq:achievable-LP} therein associated with the obfuscation scheme (the first step of the above) to incorporate the previously released $u^{([0:t-1])}$, where \eqref{eq:LP-decode} and \eqref{eq:LP-privacy} can be correspondingly modified by
\begin{equation}
\label{eq:LP-decode-general}
  p_{U^{(t)}|X^{(t)},X^{(\tau)}}(u^{(t)}|x^{(t)},x^{(\tau)}, u^{([0:t-1])}) = 0, \ \forall x^{(t)} \notin u^{(t)},
\end{equation}
and
\begin{equation}
\label{eq:LP-privacy-general}
\begin{aligned}
  & \sum_{x^{(t)}}   p(u^{(t)}|x^{(t)},x^{(\tau)}, u^{([0:t-1])}) \, p(x^{(t)}|x^{(\tau)}, u^{([0:t-1])}) \\ 
  & =  \sum_{x^{(t)}} p(u^{(t)}|x^{(t)},\tilde{x}^{(\tau)}, u^{([0:t-1])}) \, p(x^{(t)}|\tilde{x}^{(\tau)}, u^{([0:t-1])}),
\end{aligned}
\end{equation}
for any $x^{(\tau)}, \tilde{x}^{(\tau)} \in [K]$ and $u^{(t)} \in \mathscr{P}_{K}$. 
Note that the probability simplex (the third constraint in \eqref{eq:achievable-LP}) and the nonnegativity of probabilities (the fourth constraint in \eqref{eq:achievable-LP}) always have to be  guaranteed. For conciseness, we omit to write them explicitly. Hence, the optimization problem can be written by 
  \begin{equation}
\label{eq:achievable-LP-general}
    \begin{aligned}
    & \underset{\pb{u^{(t)}|x^{(t)},x^{(\tau)}, u^{([0:t-1])}}}{\text{minimize}}
    & & \mathbb{E}\left[C(N,|U^{(t)}|)\right]   & \\
    & \text{subject to}
    & & \eqref{eq:LP-decode-general}, \eqref{eq:LP-privacy-general}.
    \end{aligned}
\end{equation}

In short, the problem \eqref{eq:achievable-LP-general} can be simply modified from \eqref{eq:achievable-LP} by replacing the given distribution $p_{X^{(1)}|X^{(0)}}(x^{(1)}|x^{(0)})$ and the decision variables $p_{U|X^{(1)},X^{(0)}}(u|x^{(1)},x^{(0)})$ in \eqref{eq:achievable-LP} by $p_{X^{(t)}|X^{(\tau)}}(x^{(t)}|x^{(\tau)}, u^{([0:t-1])})$ and
$$p_{U^{(t)}|X^{(t)},X^{(\tau)}}(u^{(t)}|x^{(t)},x^{(\tau)}, u^{([0:t-1])}),$$ respectively.

As we know from Lemma~\ref{lemma:achieve}, for any given $u^{([0:t-1])}$, if $p_{X^{(t)}|X^{(\tau)}}(x^{(t)}|x^{(\tau)}, u^{([0:t-1])})$ is known, then the problem \eqref{eq:achievable-LP-general} has a feasible solution 
\[p_{U^{(t)}|X^{(t)},X^{(\tau)}}(u^{(t)}|x^{(t)},x^{(\tau)}, u^{([0:t-1])})\]
for $u^{(t)} \in \mathscr{P}_{K}$ and $x^{(t)},x^{(\tau)} \in [K]$, such that the resulting $U^{(t)}$ satisfies that 
\begin{equation*}
  \pbb{|U^{(t)}| \leq i|U^{([0:t-1])} = u^{([0:t-1])}} \geq \sum_{j=1}^{i} \theta_j(u^{([0:t-1])}),
\end{equation*}
for all $i =1,\ldots,K$, as shown in \eqref{eq:theorem-U-t}. Finally, we use the obtained $p_{U^{(t)}|X^{(t)},X^{(\tau)}}(u^{(t)}|x^{(t)},x^{(\tau)}, u^{([0:t-1])})$ to sample an obfuscation set $u^{(t)}$.

It should be noted that solving \eqref{eq:achievable-LP-general}, or more precisely obtaining the obfuscation set sampling distribution $p_{U^{(t)}|X^{(t)},X^{(\tau)}}(u^{(t)}|x^{(t)},x^{(\tau)}, u^{([0:t-1])})$ as discussed relies on knowing $p_{X^{(t)}|X^{(\tau)}}(x^{(t)}|x^{(\tau)}, u^{([0:t-1])})$. However, these quantities can not be obtained directly from the transition probabilities of the given Markov chain, since it encompasses the previously generated obfuscation sets $u^{([0:t-1])}$. As such, we need to track $p_{X^{(t)}|X^{(\tau)}}(x^{(t)}|x^{(\tau)}, u^{([0:t-1])})$ over time $t$. Roughly speaking, $p_{X^{(t)}|X^{(\tau)}}(x^{(t)}|x^{(\tau)}, u^{([0:t-1])})$ is the ``prior'' distribution at time $t$ that needs to be updated at each time $t$.

\begin{remark}  
  If $S^{(t)} = 1$, i.e., the request $X^{(t)}$ needs privacy, then we know that $\tau(t) = t$ by definition, which implies that 
  \[\theta_j(u^{([0:t-1])}) =
  \begin{cases}
    0, & j < K,\\
    1, & j = K,
  \end{cases}\]
  for any $u^{([0:t-1])}$. Referring to \eqref{eq:corollary-U-t}, it suggests that 
  \begin{equation*}
  \pbb{|U^{(t)}| = K|U^{([0:t-1])} = u^{([0:t-1])}} = 1,
\end{equation*}
which is consistent with our early observation, i.e., using a standard PIR scheme (over all $K$ messages)  if privacy is needed at time $t$. In other words, the above analysis of designing an obfuscation set $U^{(t)}$ unifies both cases $S^{(t)} = 0$ and $S^{(t)}= 1$.
\end{remark}

Now, we show how to track $p_{X^{(t)},X^{(\tau)}}(x^{(t)}, x^{(\tau)}|u^{([0:t-1])})$, that is equivalent to $p_{X^{(t)}|X^{(\tau)}}(x^{(t)}| x^{(\tau)}, u^{([0:t-1])})$ as we need, for $t = 0, 1, \ldots$. The process is essentially similar to the standard forward algorithm \cite{forward} by utilizing the Markov structure and incorporating the designed $p_{U^{(t)}|X^{(t)},X^{(\tau)}}(u^{(t)}|x^{(t)},x^{(\tau)}, u^{([0:t-1])})$.

Recall the assumption that $S^{(0)} = 1$ (implying $\tau(0) = 0$) and initial probability distribution $\pi_0$ of $X^{(0)}$ is known, so $p_{X^{(t)},X^{(\tau)}}(x^{(t)},x^{(\tau)}| u^{([0:t-1])})$ is known for $t = 0$, i.e.,
\[p_{X^{(t)},X^{(\tau)}}(x^{(t)},x^{(\tau)}| u^{([0:t-1])}) = \pi_0.\]
At each time $t = 1,2,\ldots$, we consider $S^{(t)} = 0$ or $1$ separately. 

  If $S^{(t)} = 1$, then we know that $\tau(t) = t$ by the definition of $\tau(t)$. Consider
  \begin{align*}
    & p(x^{(t+1)},x^{(\tau(t+1))}|u^{([0:t])}) \\
    & \utag{a}{=} \sum_{x^{(t)}} p(x^{(t)}|u^{([0:t])}) p(x^{(t+1)},x^{(\tau(t+1))}|x^{(t)}) \\
    & \utag{b}{=} \sum_{x^{(t)}} p(x^{(t)}, x^{(\tau(t))}|u^{([0:t-1])}) p(x^{(t+1)},x^{(\tau(t+1))}|x^{(t)}),
  \end{align*}
  where \uref{a} follows because $\tau(t+1)$ is either $t$ or $t+1$ provided that $S^{(t)} = 1$, \uref{b} follows because $U^{(t)} = [K]$ that is a constant and $\tau(t) = t$. Since $p(x^{(t+1)},x^{(\tau(t+1))}|x^{(t)})$ can be obtained straightforwardly from the transition probabilities of the Markov chain, $p(x^{(t+1)},x^{(\tau(t+1))}|u^{([0:t])})$ can be updated from $p(x^{(t)},x^{(\tau(t))}|u^{([0:t-1])})$.

  If $S^{(t)} = 0$, then we know that $\tau(t) = \tau(t-1)$ by definition. Consider 
    \begin{align*}
  & p(x^{(t+1)},x^{(\tau(t+1))}|u^{([0:t])}) \\
  & \utag{a}{=} \sum_{x^{(t)}} p(x^{(t)},x^{(\tau(t))}|u^{([0:t])}) p(x^{(t+1)},x^{(\tau(t+1))}|x^{(t)},x^{(\tau(t))}) \\
  & \propto \sum_{x^{(t)}} p(x^{(t)},x^{(\tau(t))},u^{(t)}|u^{([0:t-1])}) \\
  & ~~~~~~~~~~~~ p(x^{(t+1)},x^{(\tau(t+1))}|x^{(t)},x^{(\tau(t))}) \\
  & = \sum_{x^{(t)}} 
  p(x^{(t)},x^{(\tau(t))}|u^{([0:t-1])}) 
  p(u^{(t)}| x^{(t)},x^{(\tau(t))},u^{([0:t-1])}) \\
  & ~~~~~~~~~~~~
  p(x^{(t+1)},x^{(\tau(t+1))}|x^{(t)},x^{(\tau(t))}),
\end{align*}
where \uref{a} follows because $\tau(t+1)$ is either $\tau(t)$ or $t+1$. Since $p(u^{(t)}| x^{(t)},x^{(\tau(t))},u^{([0:t-1])})$ is the obfuscation sampling distribution by design, and $p(x^{(t+1)},x^{(\tau(t+1))}|x^{(t)},x^{(\tau(t))})$ can be directly obtained from the transition probabilities of the Markov chain, $p(x^{(t+1)},x^{(\tau(t+1))}|u^{([0:t])})$ can be updated from $p(x^{(t)},x^{(\tau(t))}|u^{([0:t-1])})$. 

Therefore, the above process keeps tracking the probability distribution $p_{X^{(t)},X^{(\tau(t))}}(x^{(t)},x^{(\tau(t))}|u^{([0:t-1])})$ for $t = 0,1,\ldots$, that is needed for the obfuscation set design.

We summarize the proposed intermittent PIR scheme as follows: at time $t$, with the known probability distribution $p_{X^{(t)},X^{(\tau(t))}}(x^{(t)},x^{(\tau(t))}|u^{([0:t-1])})$, where $u^{([0:t-1])}$ are previously generated obfuscation sets from time $0$ to $t-1$,  
\begin{enumerate}
  \item Design a sampling distribution of the  obfuscation set $U^{(t)}$, i.e.,
  \[p_{U^{(t)}|X^{(t)},X^{(\tau(t))}}(u^{(t)}|x^{(t)},x^{(\tau(t))}, u^{([0:t-1])}),\]
  such that $\pbb{X^{(t)} \in U^{(t)}} = 1$ and $U^{(t)}$ is independent of $X^{(\tau(t))}$ given the  previous obfuscation sets $u^{([0:t-1])}$, i.e.,
  \begin{equation}
  \label{eq:general-weak-privacy}
    I(X^{(\tau(t))};U^{(t)}|U^{([0:t-1])} = u^{([0:t-1])}) = 0. 
  \end{equation}
  The sampling distribution of the obfuscation set $U^{(t)}$ can be any feasible solution to the optimization problem \eqref{eq:achievable-LP-general} for known $p_{X^{(t)},X^{(\tau(t))}}(x^{(t)},x^{(\tau(t))}|u^{([0:t-1])})$, and there is an existence guarantee of a feasible solution such that the resulting $U^{(t)}$ satisfies \eqref{eq:theorem-U-t} for any given $p_{X^{(t)},X^{(\tau(t))}}(x^{(t)},x^{(\tau(t))}|u^{([0:t-1])})$.

  \item Generate an obfuscation set $u^{(t)}$ according to the designed sampling distribution 
  \[p_{U^{(t)}|X^{(t)},X^{(\tau(t))}}(u^{(t)}|x^{(t)},x^{(\tau(t))}, u^{([0:t-1])}),\]
  based on the requests $x^{(\tau(t))}$, $x^{(t)}$ and previously generated $u^{([0:t-1])}$. 

  \item Query for the request $x^{(t)}$ by using a standard PIR scheme over messages specified by $u^{(t)}$.

  \item Compute 
  \[p_{X^{(t+1)},X^{(\tau(t+1))}}(x^{(t+1)},x^{(\tau(t+1))}|u^{([0:t])})\] from the known $$p_{X^{(t)},X^{(\tau(t))}}(x^{(t)},x^{(\tau(t))}|u^{([0:t-1])}),$$ the given transition probabilities of the Markov chain, and the designed obfuscation set sampling distribution 
  \[p_{U^{(t)}|X^{(t)},X^{(\tau(t))}}(u^{(t)}|x^{(t)},x^{(\tau(t))}, u^{([0:t-1])}).\]
\end{enumerate}

Finally, let us verify that the scheme satisfies the correctness requirement and the privacy requirement (c.f. \eqref{eq:privacy}) formally. The proposed scheme guarantees that the desired message can be retrieved successful by design, since the retrieval phase is a standard PIR scheme (for the request $X^{(t)}$) over messages specified by $U^{(t)}$. From Proposition~\ref{proposition:markov}, we know that if the scheme satisfies \eqref{eq:privacy-latest} then it satisfies the desired privacy requirement in \eqref{eq:privacy}. Since the immediate privacy guarantee of the scheme is that 
\[I(X^{(\tau)};U^{(t)}|U^{([0:t-1])}) = 0,\]
which is guaranteed by \eqref{eq:general-weak-privacy} during the design, we need to show that it implies that 
$I(X^{(\tau)}; Q_{i}^{(t)}| Q_{i}^{(0)}, \ldots, Q_{i}^{(t-1)}) = 0$,
that is \eqref{eq:privacy-latest} to be justified. 
Towards this end, consider 
\begin{align*}
  & I(X^{(\tau)}; Q_{i}^{(t)}| Q_{i}^{(0)}, \ldots, Q_{i}^{(t-1)}) \\  
  & = H(X^{(\tau)}| Q_{i}^{([0:t-1])}) - H (X^{(\tau)}| Q_{i}^{(t)}, Q_{i}^{([0:t-1])}) \\
  & \utag{a}{\leq} H(X^{(\tau)}| Q_{i}^{([0:t-1])}) - H (X^{(\tau)}| U^{[0:t]}) \\
  & \utag{b}{\leq} H(X^{(\tau)}| U^{([0:t-1])}) - H (X^{(\tau)}| U^{[0:t]}) \\
  & = I(X^{(\tau)};U^{(t)}|U^{([0:t-1])}) \\
  & = 0,
\end{align*}
where \uref{a} follows because the fact that $Q_i^{(t)}$ is generated by a PIR scheme over messages in $U^{(t)}$ implies the following chain
\[X^{(\tau)} \rightarrow X^{(t)} \rightarrow U^{(t)} \rightarrow Q_{i}^{(t)},\]
and \uref{b} follows because $U^{(t)}$ is deterministic of $Q_{i}^{(t)}$ in a standard PIR scheme, i.e., knowing one of the messages in $U^{(t)}$ is being retrieved from the query $Q_{i}^{(t)}$ to the $i$-th server, which finishes the justification.

It is generally hard to obtain a closed-form formula of the download cost $\alpha^{(t)}$ at time $t$, since $\theta_j(u^{([0:t-1])})$ is not simply a function of the given transition probabilities as in the two-requests case. For this reason, we present an evaluation at the end to illustrate the download cost.

\subsubsection*{Evaluation}
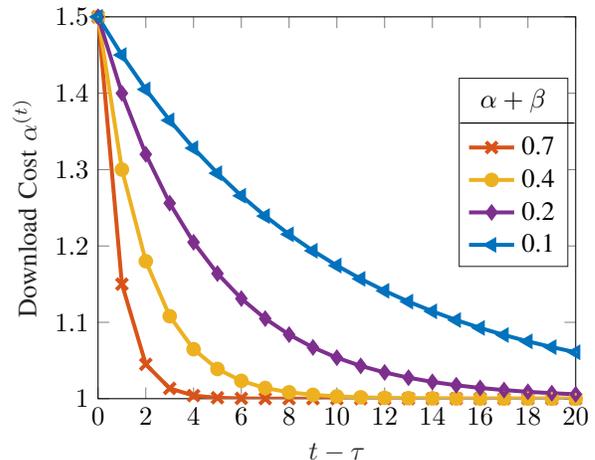
\begin{figure}[t]
    \centering
\definecolor{mycolor1}{rgb}{0.00000,0.44700,0.74100}%
\definecolor{mycolor2}{rgb}{0.85000,0.32500,0.09800}%
\definecolor{mycolor3}{rgb}{0.92900,0.69400,0.12500}%
\definecolor{mycolor4}{rgb}{0.49400,0.18400,0.55600}%
\definecolor{darkred}{rgb}{0.64, 0.0, 0.0}
%
\begin{tikzpicture}

\begin{axis}[%
width=2.5in,
height=2in,
at={(1.011in,0.642in)},
scale only axis,
xmin=0,
xmax=20,
xtick={0, 2, 4, 6, 8, 10, 12, 14, 16, 18, 20},
xlabel style={font=\color{white!15!black}},
xlabel={ $t-\tau$},
ymin=1,
ymax=1.5,
ylabel style={font=\color{white!15!black}},
ylabel={Download Cost $\alpha^{(t)}$},
axis background/.style={fill=white},
legend style={at={(0.755,0.345)}, anchor=south west, legend cell align=left, align=left, draw=white!15!black}
]

 \addlegendimage{empty legend}
 \addlegendentry{\hspace{-.55cm}\textbf{$\alpha+\beta$}}

 \addlegendimage{empty legend}
\addlegendentry{\hspace{-1.5cm} }

\addplot [color=mycolor2, line width=1.5pt, mark size=3pt, mark=x, mark options={solid, mycolor2}]
  table[row sep=crcr]{%
0 1.5\\
1 1.15\\
2 1.045\\ 
3 1.0135\\
4  1.00405\\
5 1.001215\\
6  1.0003645\\
7 1.00010935\\
8  1.000032805\\
9 1.000009842\\
10 1.000002952\\
11 1.000000886\\
12 1.000000266\\
13 1.00000008\\
14  1.000000024\\
15 1.000000007\\
16 1.000000002\\
17 1.000000001\\
18 1\\
19 1\\
20 1\\
};
\addlegendentry{0.7}

\addplot [color=mycolor3, line width=1.5pt, mark size=2pt, mark=*, mark options={solid, mycolor3}]
  table[row sep=crcr]{%
0 1.5\\
1 1.3\\
2 1.18\\
3  1.108\\
4 1.0648\\
5  1.03888\\
6 1.023328\\
7  1.0139968\\
8 1.00839808\\
9  1.005038848\\
10 1.003023309\\
11 1.001813985\\
12 1.001088391\\
13 1.000653035\\
14 1.000391821\\
15 1.000235092\\
16 1.000141055\\
17 1.000084633\\
18 1.00005078\\
19  1.000030468\\
20 1.000018281\\
};
\addlegendentry{0.4}

\addplot [color=mycolor4, line width=1.5pt, mark size=2pt, mark=diamond, mark options={solid, mycolor4}]
  table[row sep=crcr]{%
0 1.5\\
1 1.4\\
2 1.32\\
3  1.256\\
4 1.2048\\
5  1.16384\\
6 1.131072\\
7  1.1048576\\
8 1.08388608\\
9  1.067108864\\
10 1.053687091\\
11 1.042949673\\
12 1.034359738\\
13 1.027487791\\
14 1.021990233\\
15 1.017592186\\
16 1.014073749\\
17 1.011258999\\
18 1.009007199\\
19 1.007205759\\
20 1.005764608\\
};
\addlegendentry{0.2}

\addplot [color=mycolor1, line width=1.5pt, mark size=2pt, mark=triangle, mark options={solid, rotate=90, mycolor1}]
  table[row sep=crcr]{%
0 1.5\\
1 1.45\\
2  1.405\\
3 1.3645\\
4  1.32805\\
5 1.295245\\
6  1.2657205\\
7 1.23914845\\
8  1.215233605\\
9 1.193710245\\
10 1.17433922\\
11  1.156905298\\
12 1.141214768\\
13 1.127093291\\
14 1.114383962\\
15 1.102945566\\
16 1.092651009\\
17 1.083385908\\
18 1.075047318\\
19 1.067542586\\
20 1.060788327\\
};
\addlegendentry{0.1}

\end{axis}
\draw[line width=0.45pt] (7.38,5.3) -- (8.9,5.3);
\end{tikzpicture}
%
    \caption{
     The download cost $\alpha^{(t)}$ for $N = K = 2$, as a function of $t - \tau$ for different values of $\alpha+\beta$.
}
    \label{fig:th_plot}
    \vspace{-10pt}
\end{figure}

We evaluate the download cost $\alpha^{(t)}$ specified by \eqref{eq:theorem-t} and \eqref{eq:corollary-U-t} for the simplest case $N = K = 2$ as an illustration in Figure~\ref{fig:th_plot}. Similar to the previous example, suppose that the probability transition matrix of the Markov chain is 
\begin{equation*}
 P =
\begin{bmatrix}
       1- \alpha &  \alpha         \\
       \beta  & 1- \beta 
\end{bmatrix},
\end{equation*}   
such that $0 \leq \alpha, \beta \leq 1$. Since the privacy status $S^{(t)}$ affects the download cost via $\tau(t)$(c.f.\eqref{eq:define-tau}), we simulate the download cost, as a function of $t - \tau(t)$ for several values of $\alpha+\beta$, where the maximum value $1.5$ of y-axis is the download cost of a standard PIR scheme for $N = K = 2$, and the minimum value $1$ corresponds to retrieval of the desired message directly. We can observe that as $\alpha + \beta$ approaches $1$, the correlation between the requests decreases, which leads to a decrease in the download cost. As $t - \tau$ goes larger, the correlation between the current request and the latest request that needs privacy decreases, which also leads to a decrease in the download cost.

 \section{Application to location privacy}
 \label{sec:location}

As said, the reason why we are particularly interested in the Markov structure correlation is because of the motivating location privacy application. In this section, we will show how we apply the proposed intermittent PIR scheme to design an obfuscation-based location privacy protection mechanism, and discuss some specific aspects of the location privacy problem.

As we mentioned, a commonly adopted mobility model of the location trace is the Markov model \cite{shokri2011quantifying,Tandon2019location,Gunduz2021location,shokri2016privacy,chatzikokolakis2014predictive,xiao2015protecting}, i.e., the location at time (discrete time-stamp) $t$ is denoted by $X^{(t)}$ and $X^{(t)}, t = 0, 1, \ldots$ form a first-order Markov chain. Assume that each $X^{(t)}$ takes values in a common alphabet $[K]=\{1,\ldots,K\}$. 

The user may want to share his/her location with some service providers (SPs), in order to receive location-based services. In this section, we model the provided service by an information retrieval, i.e., the user sends his/her location to a SP, and then the SP responds by sending some contents according to the location. In other words, we are interested in the case such that downloading is a concern for the service quality. Also, we assume that there are multiple service providers who can provide alternative services, e.g., querying through a cloud. 

To protect the location privacy, a user may send a perturbed location to the SPs instead of the true location by sacrificing the service quality to some degree while preserving the privacy in some range. Many works \cite{oya2017back,Tandon2019location,Gunduz2021location} have been done to study the location privacy problem with different notions of privacy and utility metrics. The closest one to this paper is \cite{Tandon2019location}, where the privacy notion is information-theoretic, \Ie defined by the mutual information between true location trace and the released perturbation of locations, and the utility is defined by a non-specified distortion function.

As we keep motivating in this paper, the user may only be concerned about the privacy of some locations while others can be released without any concern about the privacy. For a time $t$, let $\mathcal{P}_{t} \subseteq [0:t]$ be the given set such that $X^{(t)}$ requires privacy if and only if $t \in \mathcal{P}$. The set $\mathcal{P}_t$ is supposed to be determined by the user, and viewed as a given parameter. The same as our discussion about intermittent PIR, the essential difference between the situation here and protecting a single location \cite{shokri2012protecting}, is that the user has to be careful when releasing the location he/she does not care about the privacy, since the location that needs privacy may be inferred due to the temporal correlation in the location trace. 

Our focus is on an extreme operational point such that privacy leakage is zero and the utility is maximized. Different from the distortion-based mechanism \cite{Tandon2019location,Gunduz2021location}, the location privacy protection mechanism in this section is obfuscation-based, \Ie mixing the true location with certain perturbed locations together and requesting the obfuscation set from SPs.

The application of our proposed intermittent private information retrieval scheme to this specific location privacy problem is straightforward, by viewing the true location $X^{(t)}$ as the request in previous sections, i.e.,
\[\text{true location} \ \xleftrightarrow{X^{(t)}} \ \text{request}.\]

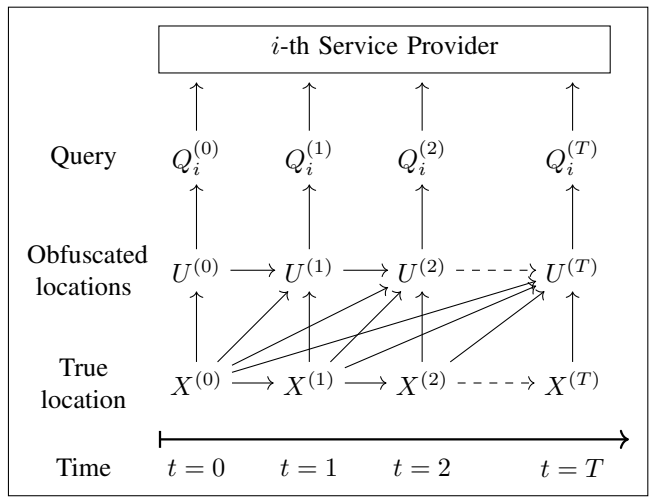
\begin{figure}[t]
  \centering
\begin{tikzpicture}[scale=0.5]
      \draw[line width=0.3mm,|->] (-1,-1.5)--(11.5,-1.5);
      \node at (-3,-2.25) {Time};

      \node[text width=1.5cm,align=center] at (-3,0) {True location};

      \node[rectangle] (x0) at (0,0) {$X^{(0)}$};
      \node at (0,-2.25) {$t=0$};

      \node[rectangle] (x1) at (3,0) {$X^{(1)}$};
      \node at (3,-2.25) {$t=1$};
      \draw[->] (x0)--(x1);

      \node[rectangle] (x2) at (6,0) {$X^{(2)}$};
      \node at (6,-2.25) {$t=2$};
      \draw[->] (x1)--(x2);

      \node[rectangle] (xT) at (10,0) {$X^{(T)}$};
      \node at (10,-2.25) {$t=T$};
      \draw[->,dashed] (x2)--(xT);

      \node[text width=1.5cm,align=center] at (-3,3) {Obfuscated locations};
      \node at (-3,6) {Query};

      \node[rectangle] (u0) at (0,3) {$U^{(0)}$};
      \node[rectangle] (y0) at (0,6) {$Q_i^{(0)}$};
      \draw[->] (u0)--(y0);

      \node[rectangle] (u1) at (3,3) {$U^{(1)}$};
      \node[rectangle] (y1) at (3,6) {$Q_i^{(1)}$};
      \draw[->] (u1)--(y1);

      \node[rectangle] (u2) at (6,3) {$U^{(2)}$};
      \node[rectangle] (y2) at (6,6) {$Q_i^{(2)}$};
      \draw[->] (u2)--(y2);

      \node[rectangle] (uT) at (10,3) {$U^{(T)}$};
      \node[rectangle] (yT) at (10,6) {$Q_i^{(T)}$};
      \draw[->] (uT)--(yT);

      \draw[->] (y0)--(0,8);
      \draw[->] (y1)--(3,8);
      \draw[->] (y2)--(6,8);
      \draw[->] (yT)--(10,8);

      \draw (-1,9.5) rectangle (11,8.25); 
      \node at (5,9) {$i$-th Service Provider};

      \draw[->] (x0)--(u0);
      \draw[->] (x0)--(u1);
      \draw[->] (x0)--(u2);
      \draw[->] (x0)--(uT);

      \draw[->] (x1)--(u1);
      \draw[->] (x1)--(u2);
      \draw[->] (x1)--(uT);

      \draw[->] (x2)--(u2);
      \draw[->] (x2)--(uT);

      \draw[->] (xT)--(uT);

      \draw[->] (u0)--(u1);
      \draw[->] (u1)--(u2);
      \draw[->,dashed] (u2)--(uT);

      \draw (-5,10) rectangle (12,-3);   
  \end{tikzpicture}
\caption{An obfuscation-based location privacy protection mechanism}
\label{fig:2}
\end{figure}

Therefore, we can directly transplant the proposed scheme in previous sections to obtain an obfuscation-based location privacy protection mechanism, as shown in Figure~\ref{fig:2}, such that at time $t$,
\begin{equation}
\label{eq:online-privacy}
  I(X^{(\mathcal{P}_{t})}; Q_{i}^{(t)}| Q_{i}^{(0)}, \ldots, Q_{i}^{(t-1)})  = 0,
\end{equation} 
where $X^{(\mathcal{P}_{t})} = \{X^{(i)}: i \in \mathcal{P} \}$, i.e., previous locations (before time $t$) that need to be protected. 

The mapping from the intermittent PIR scheme to the location privacy protection mechanism as illustrated in Figure~\ref{fig:2} should be straightforward, so we skip repeating details that can be found in previous sections. Instead, we discuss some issues regarding the privacy and utility metric in the location privacy context.

\subsubsection*{Privacy metric}
The privacy notion, i.e., left-hand side of \eqref{eq:online-privacy}, of our approach is essentially the same as the so-called \emph{online privacy} in \cite{Tandon2019location}, where the privacy is measured by   
\[\sum_{t} I(X^{(\mathcal{P}_{t})}; Q_{i}^{(t)}| Q_{i}^{(0)}, \ldots, Q_{i}^{(t-1)}),\]
i.e., the accumulation of that in \eqref{eq:online-privacy}. Since we require the stringent zero leakage, the summation over $t$ makes no difference due to the nonnegativity of the mutual information. Roughly speaking, the online privacy \eqref{eq:online-privacy} guarantees that given all previously released queries, the current query leaks zero information of all previous true locations that need privacy.

A similar privacy notion, namely \emph{offline privacy}, 
\begin{equation}
\label{eq:offline-privacy}
  I(X^{(\mathcal{P}_{T})}; Q_{i}^{(0)}, \ldots, Q_{i}^{(T)}),
\end{equation}
by assuming a finite time period $T$ for the sake of definition, was introduced in \cite{Tandon2019location}, where the authors argued that offline privacy is generally intractable to manage. 

The online privacy and the offline privacy are indeed the same as the privacy requirement \eqref{eq:privacy} and \eqref{eq:privacy-alter}
discussed in Section~\ref{sec:formulation}. As we explained, the online privacy requirement closely adheres to the causal nature of the scheme, where the  query $Q_i^{(t)}$ has to be generated at time $t$ instantly with causal information only. Also, under the stringent privacy requirement of zero leakage considered in this paper, the offline privacy metric may induce a trivial solution that the query is independent of $X^{(t)}$, which generally sacrifices the service quality too much. 

Therefore, we consider the online privacy metric in this section. However, we have to admit that the  offline privacy is theoretically interesting under a relaxed privacy requirement where privacy leakage is allowed. Indeed, \cite{Gunduz2021location} studied this notion of privacy in the framework of a Markov decision process with states $X^{(t)}$. Conceptually, the notion of offline privacy encompasses the concept of preventing the adversary from inferring  future locations, while the notion of online privacy only considers the protection of locations that have been sampled.

\subsubsection*{Utility metric}
Since we model the provided service by an information retrieval process that accommodates the obfuscation-based mechanism, the content associated with the true location can be obtained perfectly, i.e., query accuracy is perfect, by downloading more than necessary, which is different from the distortion-based mechanism that asks for the content of a perturbed location. In this sense, we consider the download cost as a utility metric to fit the obfuscation-based framework. 

We would like to slightly clarify the utility metric, as it looks different from the conventional notion, e.g., \cite{Tandon2019location,Gunduz2021location}, where the utility is measured by a single-letter distortion between the query and the true location. Since the location privacy protection mechanism is operated from another perspective in this paper, where the location privacy protection mechanism would share an obfuscated version of the true location, the quality of service is largely decided by the overhead of the content downloaded from the SPs. A motivating example here is that the user may download the map information for a larger range than he/she needs to hide the true location in some situations, e.g., augmented reality games and self-driving cars.

\section{Conclusion}
\label{sec:conclusion}
In this paper, we study the problem of intermittent private information retrieval with Markov structure correlation, where only part of the requests need privacy. 
We propose an intermittent private information retrieval scheme concatenating an obfuscation scheme and a standard PIR scheme to prevent leakage over time. The download cost is reduced compared to a standard PIR scheme, at the time when privacy is not needed. Since the Markov structure correlation is motivated by the location privacy problem, we end up by applying the proposed intermittent private information retrieval scheme to design a location privacy protection mechanism and discussing some specific issues in the location privacy problem.

\appendices
\section{Proof of Proposition~\ref{proposition:intro}}
Consider 
\begin{align*}
  & I(X^{(0)};Q^{(1)}_i) \\
  & = I(X^{(0)},Q^{(0)}_i;Q^{(1)}_i) - I(Q^{(0)}_i;Q^{(1)}_i|X^{(0)}) \\
  & = I(X^{(0)},Q^{(0)}_i;Q^{(1)}_i) \\
  & = I(Q^{(0)}_i;Q^{(1)}_i) + I(X^{(0)};Q^{(1)}_i|Q^{(0)}_i) \\
  & = I(Q^{(0)}_i;Q^{(1)}_i) + I(X^{(0)};Q^{(1)}_i,Q^{(0)}_i) - I(X^{(0)}; Q_{i}^{(0)}) \\
  & = I(Q^{(0)}_i;Q^{(1)}_i) + I(X^{(0)};Q^{(1)}_i,Q^{(0)}_i).
\end{align*}
Since 
\begin{align*}
   I(Q^{(0)}_i;Q^{(1)}_i) & \leq  I(Q^{(0)}_i;Q^{(1)}_i,X^{(0)}) \\
   & = I(Q^{(0)}_i;,X^{(0)}) + I(Q^{(0)}_i;Q^{(1)}_i|X^{(0)}) \\
   & = 0,
\end{align*}
we know that $I(Q^{(0)}_i;Q^{(1)}_i) = 0$ by the nonnegativity of the mutual information, and hence 
\[I(X^{(0)};Q^{(1)}_i) = I(X^{(0)};Q^{(1)}_i,Q^{(0)}_i),\]
which implies that $I(X^{(0)};Q_{i}^{(1)}, Q_{i}^{(0)}) = 0$ if and only if $I(X^{(0)}; Q_{i}^{(1)}) = 0$.

\section{Proof of Proposition~\ref{proposition:markov}}
The proof follows simply from the Markov structure. Consider
  \begin{align*}
    & I(X^{(\mathcal{P}_{t})}; Q_{i}^{(t)}| Q_{i}^{(0)}, \ldots, Q_{i}^{(t-1)}) \\
    & = I(X^{(\tau(t))}; Q_{i}^{(t)}| Q_{i}^{(0)}, \ldots, Q_{i}^{(t-1)})  \\
    & ~~~~~ + I\left(X^{(\mathcal{P}_{t}\backslash \tau(t))};Q_i^{(t)}|Q_{i}^{(0)}, \ldots, Q_{i}^{(t-1)}, X^{(\tau(t))} \right).
  \end{align*}
  The second term can be bounded by
    \begin{align*}
    & I\left(X^{(\mathcal{P}_{t}\backslash \tau(t))}
    ;Q_i^{(t)}|Q_{i}^{(0)}, \ldots, Q_{i}^{(t-1)}, X^{(\tau(t))} \right) \\
    & \leq I\left(X^{(\mathcal{P}_{t}\backslash \tau(t))};X^{(t)}, Q_i^{(t)}|Q_{i}^{(0)}, \ldots, Q_{i}^{(t-1)}, X^{(\tau(t))} \right) \\
    & = I\left(X^{(\mathcal{P}_{t}\backslash \tau(t))};X^{(t)}|Q_{i}^{(0)}, \ldots, Q_{i}^{(t-1)}, X^{(\tau(t))} \right) \\
    &  ~~~~~ + I\left(X^{(\mathcal{P}_{t}\backslash \tau(t))}; Q_i^{(t)}|Q_{i}^{(0)}, \ldots, Q_{i}^{(t-1)}, X^{(\tau(t))}, X^{(t)} \right) \\
    & \utag{a}{=} I\left(X^{(\mathcal{P}_{t}\backslash \tau(t))};X^{(t)}|Q_{i}^{(0)}, \ldots, Q_{i}^{(t-1)}, X^{(\tau(t))} \right) \\
    & \utag{b}{=} 0,
  \end{align*}
  where \uref{a} follows because $Q_i^{(t)}$ is a stochastic function of $X^{(\tau(t))}$, $X^{(t)}$ and $Q_{i}^{(0)}, \ldots, Q_{i}^{(t-1)}$, and \uref{b} follows because the Markov structure of $X^{(t)}$ for $t = 0,1,\ldots$ and $t \geq \tau(t) \geq \max \mathcal{P}_{t}\backslash \tau(t)$ by the definition of $\tau(t)$.

  Therefore, we obtain that 
    \begin{multline*}
    I(X^{(\mathcal{P}_{t})}; Q_{i}^{(t)}| Q_{i}^{(0)}, \ldots, Q_{i}^{(t-1)})  \\
    \leq I(X^{(\tau(t))}; Q_{i}^{(t)}| Q_{i}^{(0)}, \ldots, Q_{i}^{(t-1)}). 
  \end{multline*}
  Due to the nonnegativity of the mutual information, it is clear that 
  \begin{equation*}
  I(X^{(\tau(t))}; Q_{i}^{(t)}| Q_{i}^{(0)}, \ldots, Q_{i}^{(t-1)})  = 0, 
\end{equation*}
implies that  
\begin{equation*}
  I(X^{(\mathcal{P}_{t})}; Q_{i}^{(t)}| Q_{i}^{(0)}, \ldots, Q_{i}^{(t-1)})  = 0,
\end{equation*} 
which completes the proof.


\begin{IEEEbiographynophoto}{Fangwei~Ye}
(Member, IEEE) received the B.Eng. degree in Information Engineering from Southeast University, in 2013, and the Ph.D. degree from Department of Information Engineering, The Chinese University of Hong Kong, in 2018. From 2018 to 2020, he was a Post-Doctoral Associate with Department of Electrical and Computer Engineering, Rutgers University. He is now with the Broad Institute of MIT and Harvard.
His research interests include information theory and its applications to privacy, bioinformatics and coding opportunities in learning. 
\end{IEEEbiographynophoto}

\begin{IEEEbiographynophoto}{Salim~El~Rouayheb} (Senior Member, IEEE) received the Diploma degree in electrical engineering from the Faculty of Engineering, Lebanese University, Roumieh, Lebanon, in 2002, the M.S. degree from the American University of Beirut, Lebanon, in 2004, and the Ph.D. degree in electrical engineering from Texas A\&M University, College Station, in 2009. He is currently an Associate Professor with the ECE Department, Rutgers University, New Brunswick, NJ, USA. He was a Postdoctoral Research Fellow with UC Berkeley from 2010 to 2011, and a Research Scholar with Princeton University from 2012 to 2013. He was an Assistant Professor with the ECE Department, Illinois Institute of Technology from 2013 to 2017. His research interests are in the broad area of information theory and coding theory with applications to reliability, security, and privacy in distributed systems. He is a recipient of the NSF Career Award.
\end{IEEEbiographynophoto}

\end{document}